\documentclass[aps,showpacs,prc,nofootinbib,floatfix,]{revtex4}

\usepackage{graphicx}

\usepackage{latexsym}
\usepackage{hhline}
\usepackage{amssymb}
\usepackage{wasysym}
\usepackage{epsfig}

\sloppypar

\newcommand{\be}{\begin{equation}}
\newcommand{\ee}{\end{equation}}
\newcommand{\bea}{\begin{eqnarray}}
\newcommand{\eea}{\end{eqnarray}}
\newcommand{\ba}{\begin{array}}
\newcommand{\ea}{\end{array}}
\newcommand{\bi}{\begin{itemize}}
\newcommand{\ei}{\end{itemize}}
\newcommand{\mi}{\mbox i}

\newcommand{\refe}[1]{(\ref{#1})}

\newcommand{\mcl}{{\mathcal L}}

\newcommand{\mcp}{{\mathcal P}}

\renewcommand{\vec}[1]{\mbox{\boldmath $#1 \!\!$ \unboldmath}}

\newcommand{\smallfrac}[2]{\mbox {$\frac{#1}{#2}$}}

\newcommand{\foh}{\frac{1}{2}}

\newcommand{\fth}{\frac{3}{2}}
\newcommand{\ffh}{\frac{5}{2}}

\newcommand{\sfoh}{\smallfrac{1}{2}}

\newcommand{\sfth}{\smallfrac{3}{2}}

\begin{document}

\title{Coupled-channel analysis of the  
$\omega$-meson production in $\pi N$ and $\gamma N$ reactions
for c.m. energies up to 2 GeV.  
\footnote{Supported by FZ Juelich}}

\author{V. Shklyar\footnote{On leave from Far Eastern State University,
                       690600 Vladivostok, Russia}}
\email{shklyar@theo.physik.uni-giessen.de}
\author{H. Lenske}
\author{U. Mosel}
\author{G. Penner}
\affiliation{Institut f\"ur Theoretische Physik, Universit\"at Giessen, D-35392
Giessen, Germany}

\begin{abstract}
The pion- and photon induced reactions for the final states $\gamma N$, 
$\pi N$, $2\pi N$, $\eta N$, and $\omega N$ are studied 
within a coupled-channel effective Lagrangian approach in the energy region
from the pion threshold up to 2 GeV. To investigate the role of the nucleon 
resonances in the  different reactions we include
all known states with spin-$\foh$,-$\fth$, and -$\ffh$ and masses below 2 GeV. 
We find a strong contribution from the $D_{15}(1675)$ resonance to the $\pi N \to \omega N$
reaction. While the $F_{15}(1680)$ state only slightly influences the $\omega $ meson production
in the $\pi N$ scattering its role is enhanced in the $\omega$ photoproduction due to the
large electromagnetic coupling of this resonance. We predict the beam asymmetry $\Sigma_X$ to be 
a negative in the $\gamma p \to \omega p$ reaction near to the threshold. Above the 1.85 GeV
the  asymmetry is found  to change its sign and becomes positive at forward directions.
The presented findings can be experimentally tested at GRAAL, CLAS, and CB-ELSA facilities. 
\end{abstract}


\maketitle

\section{Introduction}
The investigation of the pion- and photon-induced reactions  on the nucleon  in the 
resonance  region  is a very interesting and  intriguing issue. 
First, the study of the pion- and  photon-nucleon reactions
provides  very interesting information on the elementary  meson-baryon dynamics which is
also  inevitable input for the investigation of  in-medium effects  in nuclear matter
either in the ground state or  at finite 
temperature. Thus, the information on the $\omega N$ elastic scattering can not be obtained 
experimentally, but can, in principle, be  extracted from an analysis of the  
$(\gamma/\pi) N \to \omega N$ data provided that all rescattering and threshold effects 
are carefully treated. This requires a dynamical coupled-channel approach  which satisfies 
the very  important condition of unitarity and is constrained by experimental data from 
all open channels. 
Secondly, the information on the  baryon resonance  spectrum can be obtained to 
distingwish between different quark model  predictions and/or lattice QCD  results. 
It is well known that some quark models predict more resonance states than discovered
so far (see \cite{Capstick:2000} and references therein). It has been assumed that these 'missing' 
resonances have small coupling to $\pi N$ and thereby are not seen in the elastic $\pi N$
scattering data. Thus, an extensive analysis of other reactions with  $\eta N$, $\omega N$,
$K \Lambda$, and $K\Sigma$ in the final state is necessary to identify properties of
those  'hidden'  resonances. 
With this aim in mind we have developed a coupled-channel effective 
Lagrangian model  
\cite{Penner:2002a,Penner:2002b,Penner:2001,Penner:PhD,Feuster:1998a,Feuster:1998b} 
that includes
the $\gamma N$, $\pi N$, $2\pi N$, $\eta N$, $\omega N$, $K \Lambda$, and $K\Sigma$ 
final states and is used for simultaneous analysis of all available experimental data in 
the energy region  $m_N+m_\pi \leqslant \sqrt{s} \leqslant $ 2 GeV. The premise is to use
the same Lagrangians for the pion- and photon-induced reactions, thereby generating 
the $u$- and $t$-  background contributions without introducing any new parameters.
In our last analysis of the pion-induced reactions \cite{shklyar:2004} it has been shown 
that while the spin-$\ffh$ states hardly influence the $\eta N$, $K \Lambda$, $K \Sigma$ final
states, the contributions from  $D_{15}(1675)$ and  $F_{15}(1680)$  to $\pi N \to \omega N$ are 
significant. However, due to the lack of hadronic data  it is not  possible to draw a firm 
conclusion about  relative resonance couplings to the $\omega N$ channel until the $\omega$ meson
photoproduction data are included \cite{Penner:2002a,Penner:2002b}.

The $\omega$ meson photoproduction is under extensive discussion in the literature because of 
the recently published high precision data from the  SAPHIR collaboration \cite{Barth:2003}.
Most of the theoretical studies of this reaction are based on the single channel 
'T-matrix' effective Lagrangian calculations  
\cite{Zhao:1998,Zhao:2000,Oh:2000,Babacan:2001,Titov:2002}.
All these  findings agree  on the importance of the $t$-channel $\pi^0$-exchange 
contributions, which has first been studied by Friman and Soyeur \cite{Friman:1995}. 
However, some  discrepancies exist between the various models on the importance of
different resonance contributions to the $\omega N$ final state. In the quark model of 
Zhao \cite{Zhao:2000} two resonance states $P_{13}(1720)$ and $F_{15}(1680)$ give 
large contributions to the $\omega$ meson photoproduction. In the approach of Titov and
Lee \cite{Titov:2002} a resonance part of the reaction is dominated by the  $D_{13}(1520)$ 
and $F_{15}(1680)$ states. An opposite observation has been made in the calculation of
Y. Oh  {\it et al.} \cite{Oh:2000}  where large contributions to the $\omega$-photoproduction 
come from the 'missing' $N\fth^+(1910)$ and  $N\fth^-(1960)$ states.
In the model of Babacan {\it et al.} \cite{Babacan:2001} resonance contributions to
$\omega N$ final state have been neglected   thereby the reaction process is described
by the only nucleon and $t$-channel production  mechanisms. A good simultaneous
description of all available experimental data on the $(\pi/\gamma) N \to \omega N$ reactions 
for $\sqrt{s}\leqslant $ 2 GeV
has been achieved in our previous study \cite{Penner:2002a,Penner:2002b}. There, strong 
resonance contributions to these reactions are found to be from the $P_{11}(1710)$ and 
$P_{13}(1900)$ states.

Since all studies predict a different individual resonance  contributions to the $\omega$
photoproduction it is interesting to look at the assumptions made in various models 
about the resonance couplings to the $ \omega N$ final state. 
In the approach of Y. Oh  {\it et al.} \cite{Oh:2000} the
$\omega NN^*$ couplings were introduced by using the quark model predictions from 
\cite{Capstick:1992,Capstick:1994}. As a result, only resonances with masses above the
$\omega N$ threshold were taken into account. In the model of Zhao \cite{Zhao:2000}
the above problem is solved by using the $SU(6)\times O(3)$ constituent quark model to also extract 
contributions from the subthreshold states. However, due to the absence of 
configuration mixing in this model the contributions from some resonances 
($S_{11}(1650)$, $D_{15}(1675)$, $D_{13}(1700)$) are strictly forbidden due to the Moorhouse 
selection rule \cite{Moorhouse:1966}. Since the experimentally 
extracted  helicity amplitude $A_{\foh}^p$ of the $S_{11}(1535)$ resonance is finite and 
large this approach was  criticized by Titov and Lee in \cite{Titov:2002}. 
To overcome this problem  these authors perform another study of the 
$\gamma p \to \omega N$ reaction \cite{Titov:2002} where a vector dominance model (VDM) 
is used to determine the $g_{\omega NN}^*$ couplings at the corresponding effective interaction
Lagrangians. Therefore, this approach considers only those resonances for which the electromagnetic 
helicity amplitudes are given in PDG \cite{pdg:2002}. Another problem with models based on the 
VDM assumption is that the $\omega NN^*$ coupling cannot be fully constrained: while the 
$A^{\omega N}_{1-\foh}$ and  $A^{\omega N}_{1+\fth}$ helicity amplitudes can be related with the
corresponding electromagnetic quantities, an additional assumptions should be put forward 
to determine $A^{\omega N}_{0+\foh}$. Therefore, in the study of \cite{Titov:2002} the $\gamma NN^*$
( $\omega NN^*$ ) dynamics has been simplified by using only one common coupling.

Assuming that some resonances might have small couplings to the $\pi N$ and $\gamma N$ final
states (see \cite{Capstick:2000,Capstick:NSTAR03}) they can only  be  excited via rescattering 
effects in other channels (e.g. $\eta N$, $\omega N$, ...). Thus, the use of a 
coupled-channel approach where all open channel are taken into account is inevitable to
identify  such resonance contributions. To our knowledge, the only calculation where the
$\omega N$ channel is treated within a coupled-channel approach is a model of Lutz {\it et al}
\cite{Lutz:2001}, where  point-like interactions are used. There, the lack
of the $J^P=\foh^+$, $J^P=\fth^+$, and $J^P=\ffh^\pm$ contributions limits the analysis 
to the near threshold region by assuming $S$-wave dominance. There is also a work by  Oh and Lee
\cite{Oh:2002sq,Oh:2002rb} where the authors  started to consider rescattering effects 
from intermediate $\pi N$ and $\rho N$ channels.

The Giessen model developed in \cite{Penner:2002a,Penner:2002b} is based on a unitary 
coupled-channel effective Lagrangian approach. It  has been successfully applied in the analysis of pion- 
and photon-induced reactions in the energy region up to 2 GeV. In this model the resonance couplings
are simultaneously constrained by available experimental data from all open channels.
Because of the complexity of the problem, our previous analysis \cite{Penner:2002a,Penner:2002b}
has been restricted to the case of resonances with spin  $J\leqslant\fth$. 
However, the contributions from the spin-$\ffh$ resonances to the final states under consideration
must be checked  explicitly, thus enlarging the model space and increasing the predictive
power of the calculations.
For example, a strong coupling of  $P_{11}(1710)$ to the
$\omega N$ has been found giving an excess structure in corresponding $\pi N$ partial wave
which is not seen in the SAID group analysis.  Since it is not a priory  clear, whether spin-$\ffh$
couplings to the $\eta N$,$\omega N$ etc. can be neglected, the calculations including all possible
contributions should be carried out in full. The motivation of this paper is to perform a new 
combined study of the $(\gamma/\pi) N $  scattering  with  $\gamma N$, $\pi N$, $2\pi N$, $\eta N$, 
and  $\omega N$ in the final state where the spin-$\ffh$ resonances are included. We check for all 
resonance contributions in the energy region up to 2 GeV. 

Our primary interest is the $\omega$ meson production. As compared to our previous
findings \cite{Penner:2002a,Penner:2002b} we expect significant changes in the resonant 
$\omega$ meson production mechanisms  by inclusion of spin-$\ffh$ resonance contributions. 
To provide an additional constraint on the resonance couplings to $\omega N$ 
we also include the recent data on the spin density matrix obtained by the SAPHIR group 
\cite{Barth:2003}.
In Sec. \ref{model} we briefly outline the main features of the applied model. 
The calculations of the $\gamma N\to \pi N$, $2\pi N$, $\eta N$ reactions and  extracted resonance 
parameters are presented in Sec.\ref{param}. The results on the $(\gamma /\pi) N \to \omega N$ 
reaction are discussed in Sec. \ref{Results}  and we finish with a Summary.

\section{The Giessen model}
\label{model}
The details of the Giessen model can be found in 
\cite{Penner:2002a,Penner:2002b,Penner:PhD,shklyar:2004}. Here we only outline the main 
features of the 
model.
The Bether-Salpeter equation (BSE) needs to be solved to obtain the scattering amplitude:
\bea
M(\sqrt{s},p,p') = K(\sqrt{s}, p, p') + i\int \frac{d^4q}{(2\pi)^4} 
V(\sqrt{s},p,q)ImG_{BS}(\sqrt{s},q)M(\sqrt{s},q,p'),\nonumber \\
K(\sqrt{s}, p, p') = V(\sqrt{s}, p, p') + \int \frac{d^4q}{(2\pi)^4} 
V(\sqrt{s},p,q)ReG_{BS}(\sqrt{s},q)M(\sqrt{s},q,p'),
\label{bse}
\eea
where the equation is split into the two constituents containing the real and imaginary parts
of the propagator $G_{BS}$.  
Here, $p$ ($k$) and $p'$ ($k'$) are the incoming and outgoing baryon (meson) four-momenta.
To date, a full solution of the equation \refe{bse} in the meson-baryon domain only exists 
for the low-energy $\pi$N scattering \cite{Lahiff:1999}, where no other channels are important.
There are many  different approximations  to the BSE which are mainly three-dimentional 
reductions (3D) of the original equation. It has been shown that there are an infinite number
of ways to perform such a reduction \cite{Yaes:1971} and there is no overwhelming reason to 
choose one particular approximation over another. Many of these approximations are intended
to avoid  singularities in the  kernel by  performing an integration over the relative energy 
in \refe{bse}  explicitly. However, due to a technical feasibility, most studies  based 
on 3D approximation are limited to elastic pion-nucleon scattering and there are only a 
few ones  \cite{Gross:1993,Surya:1996} where inelastic channels are also included. 
To solve the coupled-channel scattering problem with a large number of inelastic  channels, 
we apply the so-called K-matrix approximation where the real part of the BSE propagator
$G_{BS}$ is neglected.
This is the only way which is  feasible  for the multichannel problem and satisfies the important 
condition of unitarity. 

The imaginary part of the propagator can be written  in the form

\bea
iIm G_{BS}(\sqrt{s},q) = -i\pi^2\frac{m_{B_q} \sum_{s_B} u(p_q,s_B) \bar  u(p_q,s_B) }
{E_{B_q} E_{M_q}}\delta(k^0_q -E_{M_q})\delta(p_q^0-E_{B_q}),
\label{GBS}
\eea
thus, putting intermediate particles on their mass-shells. After the integration over the relative
energy, the  equation \refe{bse} reduces to
\bea
T^{\lambda_f\lambda_i}_{fi} = K^{\lambda_f\lambda_i}_{fi} + i\int d\Omega_n 
\sum_{n}\sum_{\lambda_n} T^{\lambda_f\lambda_n}_{fn}K^{\lambda_n\lambda_i}_{ni},
\label{GBS2}
\eea
where $T_{fi}$ is a scattering matrix and $\lambda_i$($\lambda_f$) stands for the quantum 
numbers of  initial(final) states  
\mbox{$f,i,n =$ $ \gamma N$}, $\pi N$, $2\pi N$, $\eta N$, $\omega N$, $K\Lambda$, $K\Sigma$. 
The matrix $T_{fi}$ is related to $M$ through 
$M= \smallfrac{(4\pi)^2\sqrt{s}} {\sqrt{p p' m_N m_{N'}}} T_{fi}$  \cite{Penner:2002a}.
Using the partial-wave decomposition of $T$, $K$ in terms of Wigner  functions 
(see  \cite{Penner:PhD}) the angular integration can be easily carried out and 
the equation is further  simplified to the algebraic form
\bea
T^{J\pm,I}_{fi} = \left[\frac{ K^{J\pm,I}}{1-iK^{J\pm,I}}\right]_{fi}.
\label{GBS3}
\eea

The validity of this approximation
was demonstrated by Pearce and Jennings in \cite{Pearce:1990uj} by studying different approximations
to the BSE for the $\pi N$ scattering. Considering different BSE propagators 
they concluded that an important feature of the reduced intermediate two particle
propagator is a delta function on the energy transfer. It has been argued
that there is no much difference between physical parameters obtained using the $K$-matrix 
approximation and other schemes. It has also been shown in \cite{Goudsmit:1993cp,Oset:1997it} 
that  for   $\pi N$ and $\bar K N$ scattering the  main effect from the off-shell part  is
a renormalization of couplings and masses. The assumptions made in  the $K$-matrix approach
has been checked by Sato and Lee in their calculations of the  pion photoprodution 
\cite{Sato_Lee:1996}. They find that $K$-matrix results are consistent with  
their dynamical meson-exchange model.

It should be mentioned, however, that within the $K$-matrix approach the nature of 
resonances as  three-quark excitations  or an outcome of the meson-nucleon dynamics cannot be
established. There are findings that, for example, the Roper $P_{11}(1440)$ resonance might 
be a quasibound $\sigma N$-state \cite{Krehl:1999km,Schutz:1998jx,Speth:2000zf}. There are
also studies based on the chiral model calculations where the $S_{11}(1535)$ resonance is 
dynamically  generated in the $K\Sigma$ and $\eta N$ channels \cite{Kaiser:1996js,CaroRamon:1999jf}.  
Since in the $K$-matrix approach the real part of $G_{BS}$ is neglected such resonances cannot 
appear as a quasibound state but have to be included into the potential  explicitly. Note, however, 
that a plain distinction between the three-quark and quasibound   pictures is very difficult, if
not impossible at all. Such a study may require more  extended analysis of experimental 
data (including electroproduction data) where information on the spatial content of the 
resonances can be obtained as well.

\begin{figure}
  \begin{center}
    \includegraphics[width=14cm]{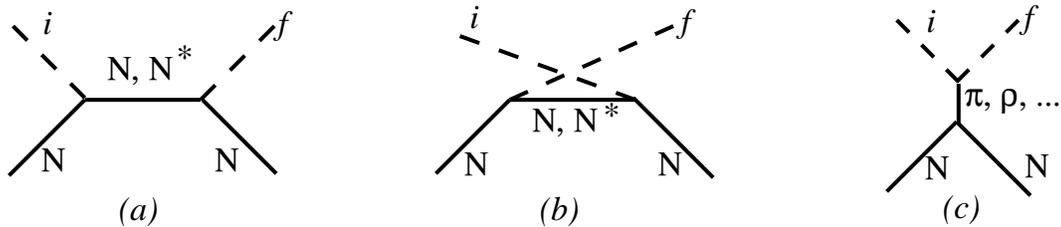}
    \caption{$s$-,$u$-, and $t$- channel contributions to the interaction potential. 
      \label{diag}}
  \end{center}
\end{figure}

Due to the smallness of the electromagnetic coupling the photoproduction reactions can be treated
perturbatively. This is equivalent to neglecting $\gamma N$ in the sum over intermediate 
states $n$ in \refe{GBS2}. Thus, for a photoproduction process the equation  \refe{GBS3} can be 
rewritten as follows
\bea
T^{J\pm,I}_{f\gamma} = K^{J\pm,I}_{f\gamma} + i \sum_{n} T^{J\pm,I}_{f n} K^{J\pm,I}_{n \gamma}. 
\label{photo}
\eea
In a similar way,  the Compton scattering amplitude can be defined as 
\bea
T^{J\pm,I}_{\gamma\gamma} = K^{J\pm,I}_{\gamma\gamma} 
+ i \sum_{n} T^{J\pm,I}_{\gamma n} K^{J\pm,I}_{n \gamma}. 
\label{Compt}
\eea 
In (\ref{photo},\ref{Compt}) summation index $n$ runs only over hadronic states. With  such
a treatment of Compton scattering problems with the gauge invariance  
during the  isospin decomposition (see \cite{Penner:2002b}) are avoided. The effects of the intermediate 
$\gamma N$ states have been checked in \cite{Penner:2002b} and found to be negligible.

\subsection{K-matrix}
The interaction potential ($K$-matrix) of the BSE is built up as a sum of 
$s$-, $u$-, and $t$-channel contributions corresponding to the tree level Feynman diagram 
shown in Fig. \refe{diag}.
Thus, the resonance and background contributions are consistently generated from the
same effective interaction Lagrangians. 
The Lagrangians used to construct the interaction potential are given in Appendix \ref{applagr} and
are discussed in \cite{Penner:2002a,Penner:2002b,Penner:PhD,shklyar:2004}. 
The $t$-channel contributions to the different final states
are summarized in  Table \ref{tchannel}.
 \begin{table}
  \begin{center}
    \begin{tabular}
      {l|c|r|r|c}
      \hhline{=====}
      & mass [GeV] & $J^P$ & $I$ & reaction \\ 
      \hhline{=====}
      $\pi$    & 0.138   & $0^-$ & $1$    & $(\gamma,\gamma),(\gamma,\pi),(\gamma,\omega)$ \\
      $\eta$   & 0.547   & $0^-$ & $0$    & $(\gamma,\gamma),(\gamma,\omega)$ \\
      $\omega$ & 0.783   & $1^-$ & $0$    & $(\gamma,\pi),(\gamma,\eta)$ \\
      \hline
      $\sigma$ & 0.650   & $0^+$ & $0$    & $(\pi,\pi)$ \\
      $f_2   $ & 1.270   & $2^+$ & $0$    & $(\pi,\pi)$ \\
      $\rho$   & 0.769   & $1^-$ & $1$    & $(\pi,\pi),(\pi,\omega),(\gamma,\pi),(\gamma,\eta)$ \\
      $a_0$    & 0.983   & $0^+$ & $1$    & $(\pi,\eta)$ \\
      \hhline{=====}
    \end{tabular}
  \end{center}
  \caption{Properties of mesons  which give contributions
    to different reactions via the $t$-channel exchange. The notation $(\gamma\gamma)$ means 
$\gamma N\to \gamma N$ etc.  
    \label{tchannel}} 
\end{table}
Using the interaction Lagrangians and values of the corresponding meson
decay widths taken from PDG \cite{pdg:2002} the following hadronic coupling constant are obtained: 
\bea
\ba{lcrclcr}
g_{\rho \pi \pi}       &=&  6.020 \; , & & 
g_{\omega \rho \pi}    &=&  2.060 \; , \\
g_{a_0 \eta \pi}       &=& -2.100 \; , & & 
g_{f_2 \pi \pi}       &=&  5.760 \; , \\
g_{\rho \pi \gamma}    &=&  0.105 \; , & & 
g_{\rho \eta \gamma}   &=& -0.928 \; , \\
g_{\omega \pi \gamma}  &=&  0.313 \; , & & 
g_{\omega \eta \gamma} &=& -0.313 \; , \\
g_{\pi \gamma \gamma}  &=&  0.037 \; , & & 
g_{\eta \gamma \gamma} &=&  0.142 \; . \\
\ea
\label{mesdeccons}
\eea
All other coupling constants are allowed to be varied during the fit.
To take into account  the finite size  of mesons and baryons each  vertex is 
dressed by a corresponding form factor:
\bea
F_p (q^2,m^2) &=& \frac{\Lambda^4}{\Lambda^4 +(q^2-m^2)^2}. 
\label{formfact} 
\eea
Here  $q$ is a c.m. four momentum of an intermediate particle and $\Lambda$ is a cutoff parameter.
 It  has been shown in \cite{Penner:2002a,Penner:2002b} that Eq. 
\refe{formfact} gives  systematically better results
 therefore we do not use any other forms for $F(q^2)$. The 
cutoffs $\Lambda$ in \refe{formfact} are treated as  free parameters being varied 
during the  calculation. However, we demand the same cutoffs in all channels for a 
given  resonance
spin  $J$ : $\Lambda^{J}_{\pi N}=\Lambda^{J}_{\pi\pi N}=\Lambda^{J}_{\eta N}=...$ etc., 
($J=\foh,~ \fth,~ \ffh$). This greatly reduces the number of free parameters; i.e. for all 
spin-$\ffh$ resonances there is  only one cutoff $\Lambda=\Lambda_{\ffh}$ for all decay channels.

The use of vertex form factors requires for a special care on preserving  
the gauge invariance when the Born contributions to photoproduction reactions 
are considered. Since the resonance and intermediate meson vertices are gauge invariant 
they can be independently  multiplied by the  corresponding form factors.  For the nucleon 
contributions to a meson photoproduction  we apply the suggestion of Davidson 
and Workman  \cite{Davidson:2001rk} and use the crossing symmetric common form factor:
\bea
\tilde F(s,u,t) = F(s) + F(u) + F(t) - F(s)F(u) - F(s)F(t)  - F(u)F(t) +  F(s)F(u)F(t).
\label{formfact2}
\eea
At present, the inelastic  $2\pi N$ channel is described  by means of an effective $\zeta N$ 
state where  $\zeta$ is an effective isovector meson with mass $m_\zeta=2m_\pi$.
We allow only resonance coupling to  $\zeta N$  therefore the decay $N^* \to \zeta N$ 
represents  a total resonance flux to the  $\rho N, \pi\Delta, \sigma N$ final states. 
To constrain  contributions to this  channel we use as an input data  
the  inelastic $2\pi N$ partial wave  cross sections extracted by Manley {\it et al.} 
\cite{Manley:1984}.  In our previous studies \cite{Penner:2002a,Penner:2002b} it has been
shown that a  good description of the $2\pi N$ channel is possible and inelastic  
data are well reproduced. Thus, in the present calculations we continue to use this 
simplified description of the $2\pi N$ channel keeping in mind that for a more reliable
description of this channel a decomposition of the $2\pi N$ final state  into 
intermediate $\rho N, \pi \Delta, \sigma N$ (similar to \cite{Manley:1992,Vrana:2000}) 
is desirable.

\subsection{$t$-channel and Born contributions}
\label{born}

\begin{table}[t]
  \begin{center}
    \begin{tabular}
      {l|r|l|r}
      \hhline{====}
      $g$ & value & $g$ & value  \\
      \hhline{====}
      $g_{NN\pi}$        &  12.85 & $g_{NN\sigma} \cdot g_{\sigma \pi \pi}$ & $ 36.01$ \\
                         &  12.85 &                                         & $ 22.92$ \\
      \hline
      $g_{NN\rho}$       &   4.40 & $\kappa_{NN\rho}$                       &   2.33 \\
                         &   4.53 &                                         &   1.47 \\
      \hline
      $g_{NN\eta}$       &   0.41 & $g_{NNa_0}$ & $-69.70$                 \\
                         &   0.10 &             & $-70.60$                  \\
      \hline
      $g_{NN\omega}$     &   4.19 & $\kappa_{NN\omega}$        &  $-0.79$ \\
                         &   3.94 &                            &  $-0.94$ \\
      \hline
      $g_{NNf_2}  $      &   5.75 &    $h_{NNf_2}$             & $-10.87$ \\
                         &   ---  &                            & ---       \\
      \hhline{====}
    \end{tabular}
  \end{center}
  \caption{Nucleon and $t$-channel couplings obtained in the present study (first line) vs.
    the results from \cite{Penner:2002a,Penner:2002b}(second line).
    \label{tabborn}}
\end{table}
The extracted $t$-channel and Born couplings are shown  in Table \ref{tabborn}. 
The obtained $g_{\pi NN}$=12.85
is slightly lower than found in other analysis, for example by SAID 
group \cite{Arndt:1995a,Arndt:1998}: 
$g_{\pi NN}$=13.13.    
Note, however, that the present calculation examines a large energy region
using only one $\pi NN$ coupling constant, thereby putting large
constraints through all production channels on this coupling and the
threshold region only plays a minor role. For example,  
the $\pi NN$coupling is especially influenced by the $t$-channel
pion exchange mechanism of the $\omega N$ photoproduction, which is due to
the restriction of using only one cutoff value $\Lambda_t$ for all
$t$-channel diagrams. 

The $\eta NN$ coupling is found to be small. This  corroborates our previous findings 
\cite{Feuster:1998a,Feuster:1998b,Penner:2002a,Penner:2002b} and the results from 
\cite{Sauermann:1997,Sauermann:1994}. Compared to \cite{Penner:2002a,Penner:2002b} also
the contribution from the $f_2(1270)$ meson exchange is taken into account.
This produces an additional background leading to a change of  the $g_{NN\sigma} \cdot g_{\sigma \pi \pi}$ 
coupling constant which appears to be larger than in the previous calculations, see Table \ref{tabborn}.

Since each interaction vertex is dressed by a form factor \refe{formfact}, special care should be 
taken  when  the values from Table \ref{tabborn} are compared  to results from other calculations.
Thus, we obtain a smaller value for the $g_{NN\omega}$=4.19 coupling constant as compared to
e.g. $g_{NN\omega}$=15.9 derived  in the Bonn model for the nucleon-nucleon scattering \cite{Machleidt:1987}.
However, it has been stressed, that taking relativistic effects into account requires the 
reduce of $g_{NN\omega}$ in the $NN$ interaction \cite{Gross:1991}. 
Moreover, in the NN scattering the $\omega NN$ coupling is utilized to describe the $t$-channel exchange
thereby its contribution is modified by a form factor. Therefore, the actual values
can be used only in combination
with the attached form factor and in the kinematical region where they have been applied to. Thus, in the
model of Titov and Lee \cite{Titov:2002} the value $g_{\pi NN}$=10.35 is used 
with the form factor of the same shape as in Eq. \refe{formfact}. However, due to the small
cutoff values $\Lambda_\omega$=0.5  applied in \cite{Titov:2002} the contribution from the
corresponding Born term is considerably suppressed.

\section{Fixing the resonance parameters}
\label{param}
In our calculations we included the  following 11
isospin $I=\foh$ resonances:
$P_{11}(1440)$,  $D_{13}(1520)$, $S_{11}(1535)$, 
$S_{11}(1650)$,  $D_{15}(1675)$, $F_{15}(1680)$, 
$P_{11}(1710)$,  $P_{13}(1720)$,  
$P_{13}(1900)$,  $F_{15}(2000)$,
and $D_{13}(1950)$, which is denoted as $D_{13}(2080)$ by 
the PDG \cite{pdg:2002}. Thus, contributions from all important resonance states in the energy 
region from the pion threshold up to 2 GeV are taken into account. 

In this energy range only $N_{17}(1990)^*$ is not included because this resonance has very large mass
close to the upper energy limit of our model. 
Thus, the contribution from this state is expected to be small.

Since the resonant part of the $\omega$ meson production amplitude is proportional to the 
two coupling constants $g_{(\gamma/\pi) NN^*}g_{\omega NN^*}$  the
resonance  couplings $g_{\gamma N N^*}$ and $g_{\pi N N^*}$ are  needed  to be  fixed first.
Taking the best hadronic result from \cite{shklyar:2004} we perform a new coupled-channel 
calculation of the  pion- and photon-induced reactions in the region up to 2 GeV  where
free coupling constants are constrained by full set of experimental data 
in the $\gamma N$, $\pi N$, $2\pi N$, $\eta N$, $K\Lambda$, $K\Sigma$, and $\omega N$ channels.
We obtain a significantly improved $\chi^2$ for the photon induced reactions with  $\omega N$, $K\Lambda$, 
and $K\Sigma$ in the final states: $\chi^2$=4.2(6.25), 2.1(3.95), 1.6(2.74) respectively,
where the values from our previous results are shown in brackets. For other channels the resulting 
$\chi^2$ are very similar to the values from the best global fit in \cite{Penner:2002a,Penner:2002b}.
As pointed out before, in this paper we concentrate on the $\omega$ meson production.  
The results on the associated strangeness production are presented in \cite{shklyar:preparation}.
First, we briefly discuss  the results on the $\pi N$ and $2\pi N$  
channels.

\begin{figure}
  \begin{center}
    \parbox{17cm}{
      \parbox{17cm}{\includegraphics*[width=14cm]{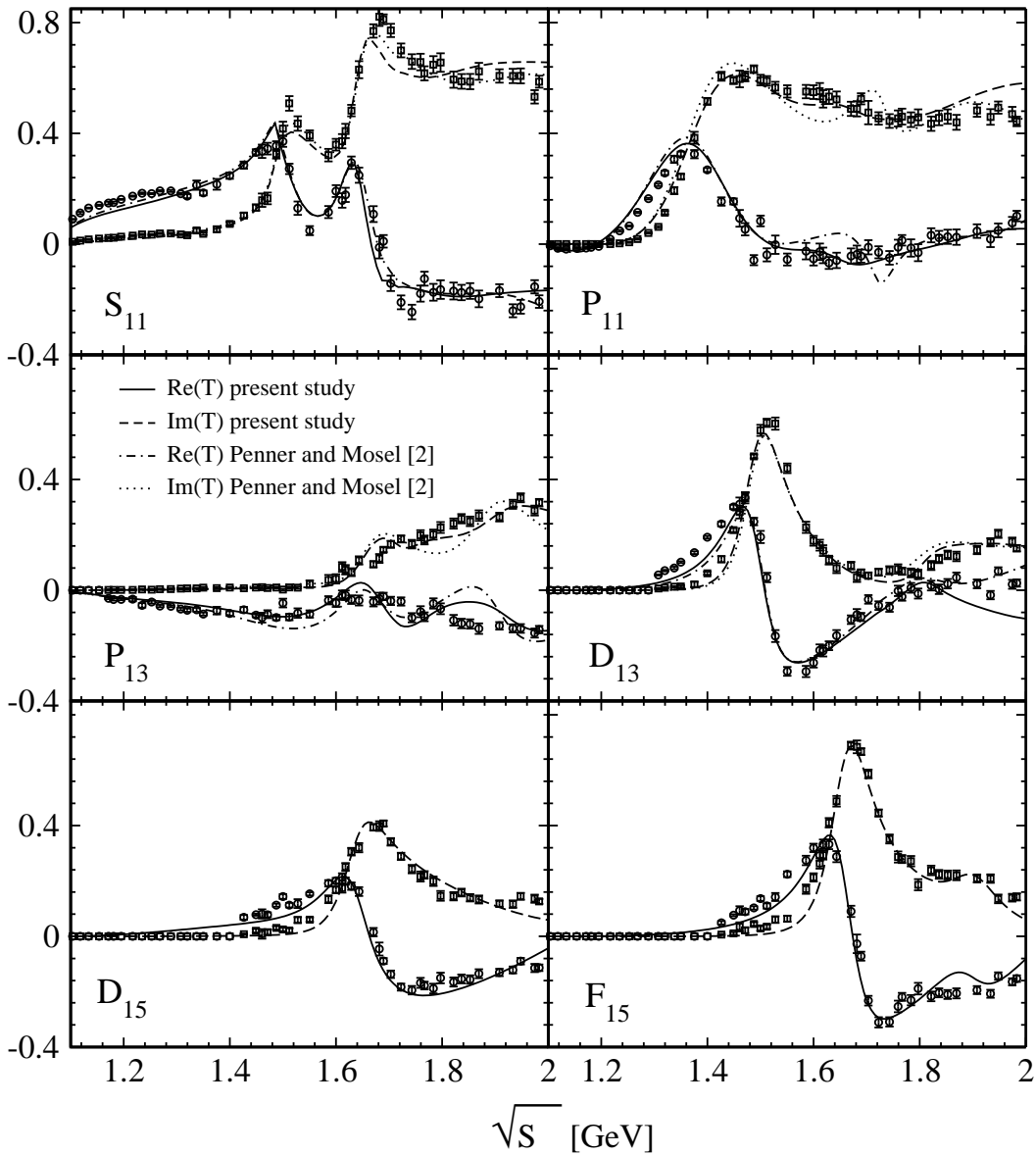}}
       }
       \caption{
$\pi N \to \pi N$ elastic partial waves for $I$=$\foh$. The solid (dashed) lines corresponds 
to the real (imaginary) part of the amplitude. Our previous best global results from 
\cite{Penner:2002a}  are shown by the dash-dotted and dotted lines. 
The data are taken from the SAID analysis \cite{Arndt:2003}.
      \label{piN_I12}} 
  \end{center}
\end{figure}

\begin{figure}
  \begin{center}
    \parbox{17cm}{
      \parbox{17cm}{\includegraphics*[width=14cm]{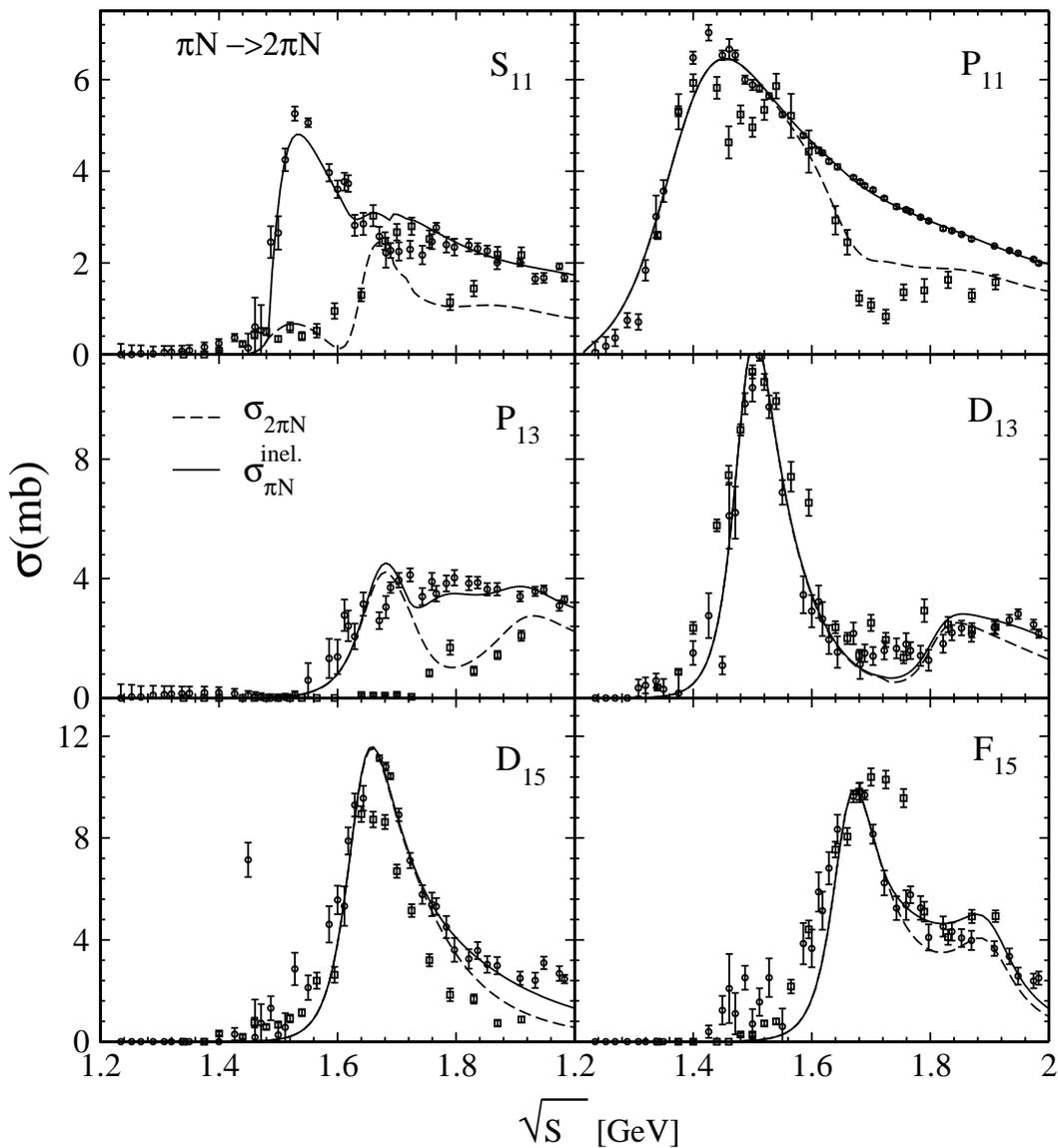}}
       }
       \caption{
$\pi N \to 2\pi N$ total partial wave cross sections and $\pi N$ inelasticities for $I$=$\foh$. 
The solid (dashed) lines 
corresponds to $\sigma^{inel.}_{\pi N}$ ($\sigma_{2\pi N}$).  The data are taken from \cite{Manley:1984,Arndt:2003}.
      \label{2piN_I12}} 
  \end{center}
\end{figure}

\begin{figure}
  \begin{center}
    \parbox{17cm}{
      \parbox{17cm}{\includegraphics*[width=15cm]{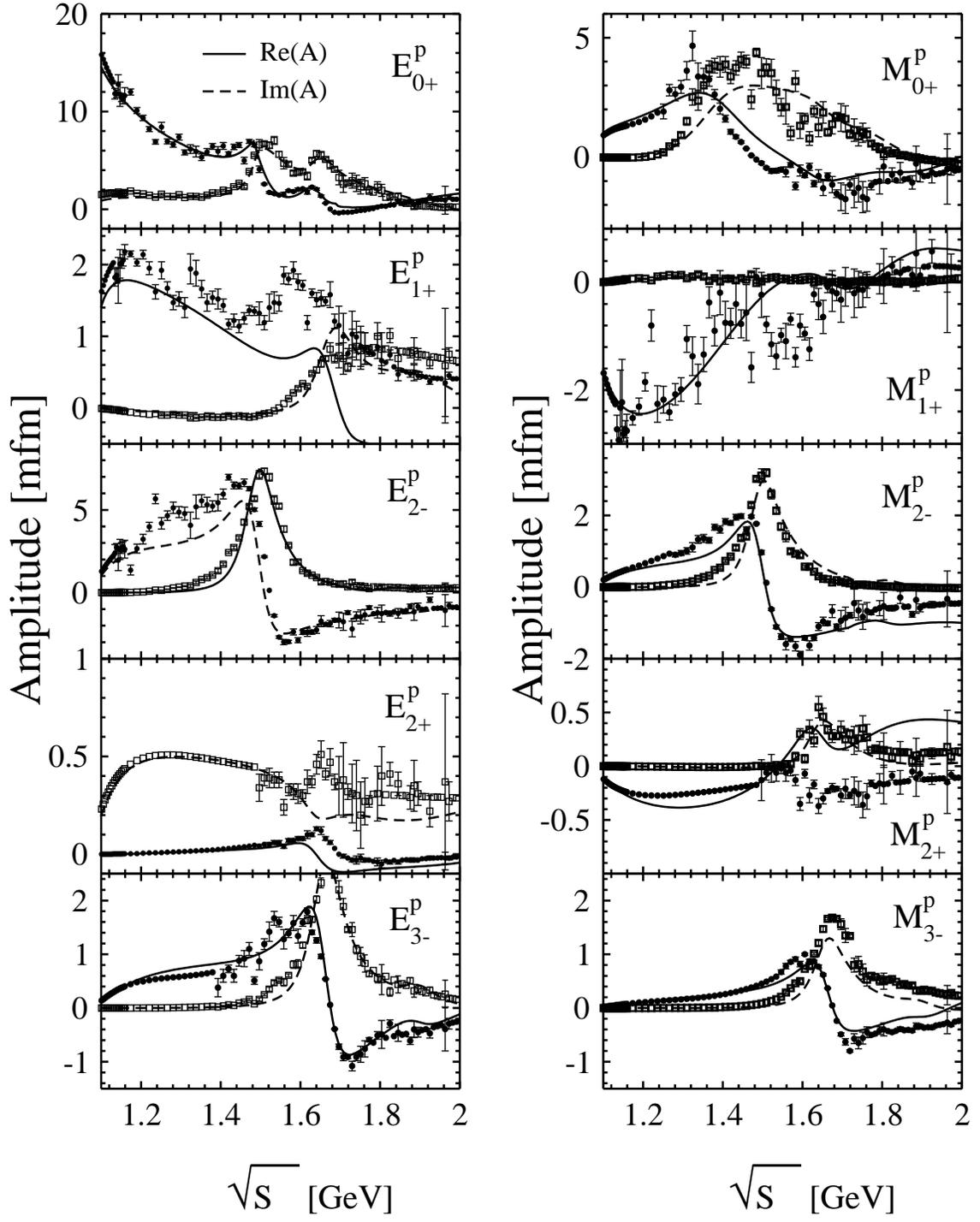}}
       }
       \caption{
$\gamma N \to \pi N$ proton multipoles. The solid (dashed) line 
corresponds to the calculated real (imaginary) part of the amplitude.
The SAID(SM01) data are taken  from \cite{Arndt:1996}.
      \label{gN_I12p}} 
  \end{center}
\end{figure}
\begin{figure}
  \begin{center}
    \parbox{17cm}{
      \parbox{17cm}{\includegraphics*[width=15cm]{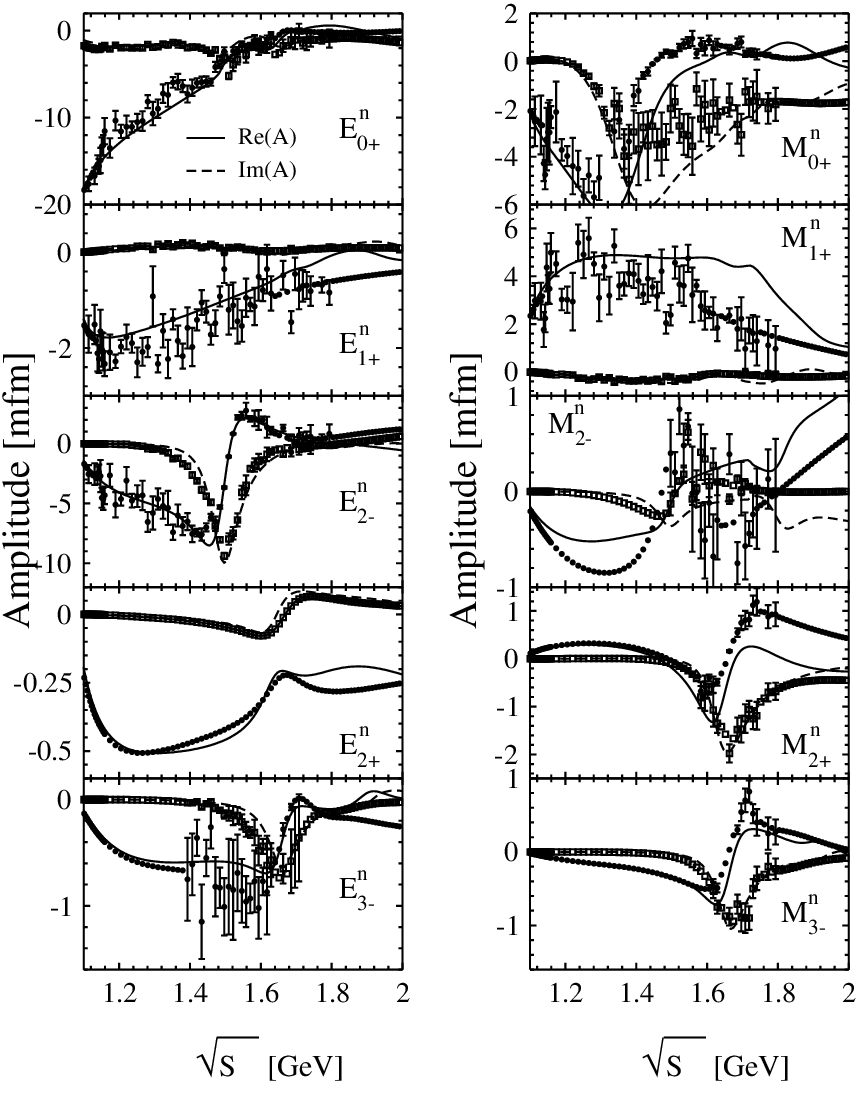}}
       }
       \caption{
$\gamma N \to \pi N$ neutron multipoles. Notation  as in Fig. \refe{gN_I12p}.
      \label{gN_I12n}} 
  \end{center}
\end{figure}

\subsection{$\pi N$ final state}
\label{piN}

The results for the elastic $\pi N$ partial wave amplitudes with isospin $I=\foh$ in 
comparison with our previous findings from \cite{Penner:2002a} are shown in Fig. \refe{piN_I12}.
The  calculated total $2\pi N$ partial wave cross sections and corresponding
$\pi N$ inelasticities are presented in Fig. \refe{2piN_I12}.  Note, that the $\pi N$
inelasticities are not fitted but obtained as a  sum of the individual contributions from
all open channels.

Since the SAID multipole data have  rather small error bars and but  scatter a lot
the description of the  pion-photoproduction multipoles turns out be a difficult task. 
The calculated multipoles
are shown in Figs. (\ref{gN_I12p},\ref{gN_I12n}) in comparison with the energy-independent 
SAID solution \cite{Arndt:1996}. For those energy region where the single-energy results were 
absent, the gaps were filled by the energy-dependent solution of the SAID group.

There  are two resonances $S_{11}(1535)$ and $S_{11}(1650)$
which  are necessary to describe the $S_{11}$ partial wave.
In order to describe the second resonance peak at $\sqrt{s}=$1650 MeV in the $E_{0+}^{p}$ multipole  
the fit shifts the mass of $S_{11}(1650)$ to the lower value 1661 MeV.
This leads to a somewhat worse description of the $S_{11}$ partial wave  in the second resonance 
region, see Fig. \refe{piN_I12}. In the  analyses  \cite{Vrana:2000,Batinic:1995,Manley:1992}  
a third resonance $S_{11}(2090)$ has been identified below 2 GeV. Moreover, in the model of
Chen et al. \cite{Chen:2002}, a fourth $S_{11}$ resonance has been found above 2 GeV. 
We have also checked whether the  inclusion of a third resonance would improve  the results. 
However, the fit gives zero width for this resonance thereby we do not find any support for
this state in the present calculations.

\begin{table}[t]
  \begin{center}
    \begin{tabular}
      {r|r|r|r|r|r|r|r}
      \hhline{========}
      $\Lambda_N$ [GeV] & 
      $\Lambda_\foh^h$ [GeV] & 
      $\Lambda_\fth^h$ [GeV] & 
      $\Lambda_\ffh^h$ [GeV] & 
      $\Lambda_\foh^\gamma$ [GeV] & 
      $\Lambda_\fth^\gamma$ [GeV] & 
      $\Lambda_\ffh^\gamma$ [GeV] & 
      $\Lambda_t^{h,\gamma}$ [GeV] \\
      \hhline{========}
  0.952 & 3.80 & 0.970 & 1.13 &  1.67  &  4.20  &  1.167  &  0.77 \\
  0.960 & 4.30 & 0.960 & ---  &  1.69  &  4.30   &   ---    & 0.70 \\
      \hhline{========}
    \end{tabular}
  \end{center}
  \caption{Cutoff values for the form factors (first line) in 
    comparison with the previous global results from \cite{Penner:2002b}
    (second line). The lower index denotes the intermediate particle, i.e. 
    $N$: nucleon, $\foh$: spin-$\foh$ resonance, $\fth$: spin-$\fth$, $\ffh$: 
    spin-$\ffh$ resonance, $t$: $t$-channel meson. The upper index $h$($\gamma$)
    denotes whether  the value is applied to a hadronic or elctromagnetic vertex.
    \label{tabcutoff}}
\end{table}

The inclusion of spin-$\ffh$ resonance contributions 
greatly changes the $\omega$ meson production mechanism. 
Through  the coupled-channel 
effects, the change in the $\omega N$ channel  affects other  reactions
which is also seen  in the 
$\pi N$ partial waves, Fig. \refe{piN_I12}.  In the present  study  we find   minor contributions
from the $P_{11}$  resonances to the $\omega N$ final state thereby a kink structure at 
$\sqrt{s}=$1.72 GeV is not visible any more in the $P_{11}$ partial wave, see Section \ref{Results}.
This is in line with the results of the SAID analysis \cite{Arndt:2003} which show an almost flat
behaviour in this  energy region. There are only minor changings in other partial waves 
as compared to our previous calculations. 

For the mass and width of the Roper resonance we find $M=1517$ MeV and $\Gamma=608$ MeV which turn out
to be rather large in comparison with results from other studies, see Table \ref{tab12}. 
However, the baryon analysis of
Vrana et al. \cite{Vrana:2000} give 490$\pm$120 MeV for the  total width. There are  also 
calculations of Cutkosky and Wang \cite{Cutkosky:1990} where a width of 661 and 545 MeV 
have been extracted from the analysis of the $\pi N$ and $2\pi N$ data. The properties of the 
$P_{11}(1440)$ are found to be very sensitive to the background contributions, i.e. to the
interference pattern between nucleon and the $t$-channel $\rho$-meson exchange. 
Since the description of the $E_{0+}^{p/n}$ multipole requires a rather soft nucleon 
cutoff (see Table \refe{tabcutoff}),  the
description of the $S_{11}$ and $P_{11}$ wave becomes worse. The fit tried to compensate 
this effect by enlarging the mass and width of $P_{11}(1440)$. Note, however, that the
$\pi N$ and $2\pi N$ branching rations of $P_{11}(1440)$ are found to be consistent with the 
result from other analysis, see Table \ref{tab12}. 
We find a  second state $P_{11}(1710)$ which is completely 
inelastic and has a very small branching ratio to $R_{\pi N}$ as required by the SAID data.
However, the decrease in $\pi N$ coupling of this resonance is compensated by the increase
of the  $R_{2\pi N}$ and $R_{\eta N}$ keeping the production   $R_{\pi N} \cdot R_{2\pi N}$ and 
$R_{\pi N} \cdot R_{\eta N}$ in line with the results from our previous global fit 
\cite{Penner:2002a,Penner:2002b}.
Hence a  good description of the $2\pi N$ cross 
section and $\pi N$ inelasticity is possible, see Fig. \refe{2piN_I12}.
Note, that the vanishing
$\pi N$ decay width of this resonance is also found by the SAID group \cite{Arndt:2003}.
 
The  $P_{13}$ inelasticity from the SAID analysis \cite{Arndt:2003} in the energy region
between 1.55 and 1.7 GeV  increases up to 4 mb while the $2\pi N$ cross section extracted
by Manley et al. \cite{Manley:1984} is found to be zero, see Fig. \refe{2piN_I12}.
This might be an indication that
either the extracted $2\pi N$ cross section is inconsistent with the SAID data or  another 
inelastic channel (not $2\pi N$) gives noticeable contributions to this partial wave. 
The same problem has also been reported by Manley and Saleski in their combined analysis 
of the $\pi N\to \pi N$ and $\pi N\to 2\pi N$ reactions \cite{Manley:1992}. These authors
suggested that the discrepancy between the data can be related with the inelastic 
contributions from the $3\pi N$ final state. So far, no analysis has been made to describe, 
e.g., the $\rho \Delta$ channel. Therefore, we follow \cite{Manley:1992} and increase the error bars  
of the original $2\pi N$ data to prevent the calculations from putting much weight to this
discrepancy.

\begin{table}[t]
  \begin{center}
    \begin{tabular}
      {l|l|l|l|l|l|c||c |c|c }
      \hhline{==========}
      $L_{2I,2S}$ & mass & $\Gamma_{tot}$ &
      $R_{\pi N}$ & $R_{2\pi N}$ & $R_{\eta N}$ &
      $R_{\omega N}$ & $g^1_{RN\omega}$ & $g^2_{RN\omega}$ & $g^3_{RN\omega}$ \\
      \hhline{==========}
      $S_{11}(1535)$ 
& 1526    & 136     & 34.4  & $ 9.5(+)$ & $56.1(+)$  & ---  & $  3.79$ & $  6.50$ & --- \\
& 1534(7) & 151(27) & 51(5) &           &            &      &            &            &     \\
& 1542(3) & 112(19) & 35(8) &           & 51(5)      &      &            &            &     \\
      \hline
      $S_{11}(1650)$ 
& 1664    & 131     & 72.4  & $23.1(+)$ & $ 1.4(-)$  & ---  & $ -1.13$   & $ -3.27$   & --- \\
& 1659(9) & 173(12) & 89(7) &           &            &      &            &            &     \\
& 1689(12)& 202(40) & 74(2) &           & 6(1)       &      &            &            &     \\
      \hhline{==========}
      $P_{11}(1440)$ 
& 1517    & 608     & 56.0  & $44.0(+)$ & $ 2.82^a$  & ---  & $  1.53$   & $ -4.35$   & --- \\
& 1462(10)& 391(34) & 69(3) &           &            &      &            &            &     \\
& 1479(80)& 490(120)& 72(5) &           & 0(1)       &      &            &            &     \\
      \hline
      $P_{11}(1710)$ 
& 1723    & 408     &  1.7  & $49.8(-)$ & $43.0(+)$  &  0.2 & $ -1.05$   & $ 10.5   $ & --- \\
& 1717(28)& 480(230)& 9(4)  &           &            &      &            &            &     \\
& 1699(65)& 143(100)& 27(13)&           & 6(1)       &      &            &            &     \\
      \hhline{==========}
      $P_{13}(1720)$ 
& 1700    & 152     & 17.1  & $78.7(+)$ & $ 0.2(+)$  & ---  & $ -6.82  $ & $ -5.84  $ & $ -8.63  $ \\
& 1717 (31)& 380(180)& 13(5)&           &            &      &            &            &     \\
& 1716(112)& 121(39)&$\;\,$5(5)&        & 4(1)       &      &            &            &     \\

      \hline
      $P_{13}(1900)$ 
& 1998    & 404     & 22.2  & $59.4(-)$ & $ 2.5(-)$  & 14.9 & $ 5.8    $ & $ 14.8   $ & $-9.9     $ \\
& 1879(17)& 498(78) & 26(6) &           &            &      &            &            &     \\
& NF      &         &       &           &            &      &            &            &     \\
      \hhline{==========}
      $D_{13}(1520)$ 
& 1505    & 100     & 56.6  & $43.4(-)$ & $ 1.2^b(+)$& ---  & $  3.35  $ & $  4.80  $ & $ -9.99  $ \\ 
& 1524(4) & 124(8)  & 59(3) &           &            &      &            &            &     \\
& 1518(3) & 124(4)  & 63(2) &           & 0(1)       &      &            &            &     \\
      \hline
      $D_{13}(1950)$ 
& 1934    & 859     & 10.5  & $68.7(+)$ & $ 0.5(-)$  & 20.1 &   $-10.5$  &  $-0.6$ &  $ 17.4  $ \\
& 1804(55)&450(185) & 23(3) &           &            &      &            &            &     \\
& 2003(18)&1070(858)& 13(3) &           & 0(2)       &      &            &            &     \\
      \hhline{==========}
      $D_{15}(1675)$ 
& 1666    & 148     & 41.1   & $58.5(+)$& $ 0.3(+)$  & ---  & $109$      & $-99.00$ & $ 83.5$ \\
& 1676(2) & 159(7)  & 47(2) &           &            &      &            &            &     \\
& 1685(4) & 131(10) & 35(1) &           & 0(1)       &      &            &            &     \\
      \hhline{==========}
      $F_{15}(1680)$ 
& 1676    & 115     & 68.3  & $31.6(+)$ & $ 0.0(+)$  & ---  &   $12.40 $ &  $-35.99$  & $-78.28$ \\
& 1684(4) & 139(8)  & 70(3) &           &            &      &            &            &     \\
& 1679(3) & 128(9)  & 69(2) &           &  0(1)      &      &            &            &     \\
      \hline
      $F_{15}(2000)$ 
& 1946    & 198     &  9.9  & $87.2(-)$ & $ 2.0(-)$  &  0.4 & $-19.6$    & $ 19.3$    & $ 23.14$ \\
& 1903(87)& 490(310)&  8(5) &           &            &      &            &            &     \\
& NF      &         &       &           &            &      &            &            &     \\
      \hhline{==========}
    \end{tabular}
  \end{center}
  \caption{Properties of $I=\foh$ resonances extracted in the
    present study  (1st line) in comparison with the
    values from \cite{Manley:1992} (2rd line), and 
    \cite{Vrana:2000} (3th line). In brackets, the estimated
    errors are given. The mass and total width are given in MeV, the
    decay ratios in percent. 
    $^b$: The decay ratio is given in 0.1\permil.
    \label{tab12}} 
\end{table}

\begin{table}[t]
  \begin{center}
    \begin{tabular}
      {l|ll|ll }
      \hhline{=====}
      $L_{2I,2S}$ & $A^{p}_{\foh}$ &
      $A^{n}_{\foh}$ & $A^{p}_{\fth}$ & $A^{n}_{\fth}$ \\
      \hhline{=====}
      $S_{11}(1535)$ 
    &   92      &  -13     &  \multicolumn{2}{c}{---}  \\
    &   90(30)  &  -46(27) &  \multicolumn{2}{c}{---}  \\
    &   60(15)  &  -20(35) &   \multicolumn{2}{c}{---} \\
\hline
$S_{11}(1650)$ 
    &   57      &  -25     &  \multicolumn{2}{c}{---} \\
    &   53(16)  &  -15(21) & \multicolumn{2}{c}{---}  \\
    &   69(5)   &  -15(5)  &  \multicolumn{2}{c}{---} \\
\hhline{=====}
$P_{11}(1440)$ 
    &  -84      &  138     & \multicolumn{2}{c}{---}  \\
    &  -65(4)   &  40(10)  & \multicolumn{2}{c}{---}  \\
    &  -63(5)   &  45(15)  & \multicolumn{2}{c}{---}  \\
\hline
$P_{11}(1710)$ 
    &    -50    &   68     & \multicolumn{2}{c}{---}  \\
    &    9(22)  & -2(14)   & \multicolumn{2}{c}{---}  \\
    &    7(15)  & -2(15)   & \multicolumn{2}{c}{---}  \\
\hhline{=====}
$P_{13}(1720)$ 
     &  -65     &    1     &   35      & -4      \\
     &   18(30) &  1(15)   & -19(20)   & -29(61) \\
     &  -15(15) &  7(15)   &  7(10)    & -5(25)  \\
\hline
$P_{13}(1900)$ 
    &   -8      &  -19     &    0      &   6 \\
    &   NG      &          &           &     \\
\hhline{=====}
$D_{13}(1520)$ 
    &  -13      &   -70    &  145      & -141       \\
    &  -24(9)   & -59(9)   &  166(5)   & -139(11)   \\
    &  -38(3)   & -48(8)   &  147(10)  & -140(10)   \\
\hline
$D_{13}(1950)$ 
    &   11      &       40 &        26 &  -33  \\
    &   NG      &          &           &       \\
\hhline{=====}
$D_{15}(1675)$ 
    &    9      &  -56     &    20     &  -84      \\
    & 19(8)     & -43(12 ) &  15(8)    & -58(13)   \\
    & 15(10)    & -49(10 ) &  10(7)    & -51(10)   \\
\hhline{=====}
$F_{15}(1680)$ 
    &   3     &   30       &   115     & -48       \\
    & -15(6)  &   29(10)   &   133(12) & -33(9)    \\
    & -10(4)  &   30(5)    &   145(5)  & -40(15)   \\
\hline
$F_{15}(2000)$ 
    &    10   &  9    &  25       &  -4   \\
    &    NG   &       &           &     \\
      \hhline{=====}
    \end{tabular}
  \end{center}
  \caption{Helicity decay amplitudes  of $I=\foh$ resonances
             ( in $10^{-3}$ GeV$^{-\foh}$)
           considered in the present study (first line);
          second line: values from PDG \cite{pdg:2002};
          third line: results of SAID group \cite{Arndt:1996};
          "NG": not given. 
    \label{helict}} 
\end{table}

There are two resonances $P_{13}(1720)$ and $P_{13}(1900)$ which contribute to 
the $P_{13}$ partial wave. The properties of the first resonance are not well fixed:
Manley and Saleski \cite{Manley:1992} give for the total width 383$\pm$179 MeV while in 
the analysis of Vrana et al. \cite{Vrana:2000} the another value of 121$\pm$39 MeV 
has been  extracted. We obtain
$\Gamma=152$ MeV which is close to the results of \cite{Vrana:2000}. The second resonance 
$P_{13}(1900)$ is rated by PDG by two stars and was only  found in the calculations
of Manley and Saleski \cite{Manley:1992}. We also find a necessity of the inclusion of this
resonance to describe the $P_{13}$ partial wave data, see Fig. \refe{piN_I12}. However, 
only a satisfactory description of the real 
part of the $E^p_{1+}$ multipole in the  energy region between 1.5 and 2 MeV is still 
possible, see Fig. \refe{gN_I12p}. This problem is due to a missing  background 
contributions to $E^p_{1+}$. Since the problem starts at the same energy where the 
discrepancy between the SAID inelasticity and the $2\pi N$ cross section in the 
$P_{13}$ partial wave is observed, it might be also related to the lack of the 
$3\pi N$ contributions to this channel \cite{Penner:2002b} as discussed above. 
Therefore, it would be desirable to account for $3\pi N$ contributions  in 
future investigations.

The mass and width of $D_{13}(1520)$ extracted in the present calculations are close to values 
obtained by  Arndt et al. \cite{Arndt:1996,Arndt:1995a}: $1516\pm 10$ and  $106\pm 6$ 
MeV correspondingly. Manley and Saleski \cite{Manley:1992} and Vrana et al. 
\cite{Vrana:2000} give somewhat larger values, see Table \ref{tab12}. 
Note, however, that the calculated
$\pi N$ and $2\pi N$ branching ratios are very close to that of \cite{Vrana:2000,Manley:1992}.
Apart from the well-established resonance $D_{13}(1520)$  we also include  the second state 
$D_{13}(1950)$ which is denoted as $D_{13}(2080)$ by the PDG \cite{pdg:2002}. This resonance 
is poorly established: in all calculations it appears to be almost inelastic and weakly coupled 
to $\pi N$.  Despite on its small decay ration to $\pi N$, this 
resonance turns out to be important due to rescattering effects. Without this state the
calculations result in considerably worse $\chi^2$.
The $D_{13}(1950)$ state is found to be rather broad in the present calculations: the obtained
width is about 860 MeV and mass 1934 MeV. Other baryon analysis also identify this state with
a  large width: Vrana et al. \cite{Vrana:2000} find $\Gamma=1070\pm 858$ MeV and Manley and Saleski 
\cite{Manley:1992} obtain $\Gamma=447\pm 185$ MeV. Note, that we do not  find any indication for the
$D_{13}(1700)$ resonance contribution in the energy region between 1.7 and 1.9 GeV as compared to 
results of \cite{Vrana:2000,Manley:1992}. 
In all calculations the fit gives almost zero width for this resonance, hence its contributions vanish.

There is a clear resonance peak in the $D_{15}$ partial wave, (see Fig. \refe{piN_I12}) which 
corresponds to  the $D_{15}(1675)$ resonance. The comparison of the $2\pi N$ total cross section
extracted by Manley et al. \cite{Manley:1984} with the SAID inelasticity  shown in Fig. \refe{2piN_I12}
reveals the missing inelastic flux of 2 mb around 1.65 GeV. It has been shown in \cite{shklyar:2004}
that this flux cannot be   absorbed by neither the $\eta N$, $K\Lambda$, or 
$K\Sigma$ channels. Thus, we conclude that either the 
$\pi N$ and $2\pi N$ data are inconsistent with each other or other open channels 
(e.g. 3$\pi N$) must be taken into account. To overcome this problem and to describe the 
$\pi N$ and $2\pi N$ data in the $D_{15}$  partial wave the error bars of the original $2\pi N$ data  
\cite{Manley:1984}  were weighted by a factor 3. The same procedure was also used by 
Vrana et al. \cite{Vrana:2000} and Cutkosky et al. \cite{Cutkosky:1990} to fit the inelastic data.
We find an important subthreshold contributions from the $D_{15}(1675)$ resonance  to 
the $\pi N \to \omega N$ reaction, see Section \ref{Results}. Hence, the $D_{15}$  inelastic 
contribution of about 1 mb shifts from the $2\pi N$ channel to $\omega N$ above 1.8 GeV as shown in 
Fig. \refe{2piN_I12}. 

Apart from the well established $F_{15}(1680)$ we also find an indication for the second $F_{15}(2000)$
resonance to describe the hight energy tail of the $F_{15}$ partial wave amplitude, 
as seen in  Fig. \refe{piN_I12} by the shoulder around 1950 MeV. 
The evidence for this state was  also found in earlier  works \cite{Manley:1992,Hohler:1979yr}. 
A visible inconsistency  between the inelastic SAID data and  the 
$2\pi N$ cross section from \cite{Manley:1984} above 1.7 GeV  can  be seen in $F_{15}$ wave,
see  in Fig. \ref{2piN_I12}.  The three  data points at 1.7, 1.725, and 1.755 GeV  are, therefore, 
excluded from the fitting procedure. 
 
The parameters of the  $D_{15}(1675)$ and $F_{15}(1680)$ resonances are in line with the results from
other groups \cite{Vrana:2000,Manley:1992,pdg:2002}.  The properties of the  $F_{15}(2000)$
state  differ strongly in the various analyses: Manley and Saleski \cite{Manley:1992} give   
$490\pm 310$ MeV for the total decay width while other studies \cite{Arndt:1995a,Hohler:1979yr} find 
it at the level of $95\div 170$ MeV. Moreover, this state has not been identified in the  investigations 
of \cite{Vrana:2000,Batinic:1995}. Since the $F_{15}(2000)$ resonance is found to be strongly 
inelastic with  84-88\% of inelasticity absorbed by the $2\pi N$ channel, more  $2\pi N$ 
data above 1.8 GeV (cf. Fig. \ref{2piN_I12}) are needed for a reliable determination of the properties 
of this state.

\begin{figure}
  \begin{center}
    \parbox{17cm}{
      \parbox{17cm}{\includegraphics*[width=17cm]{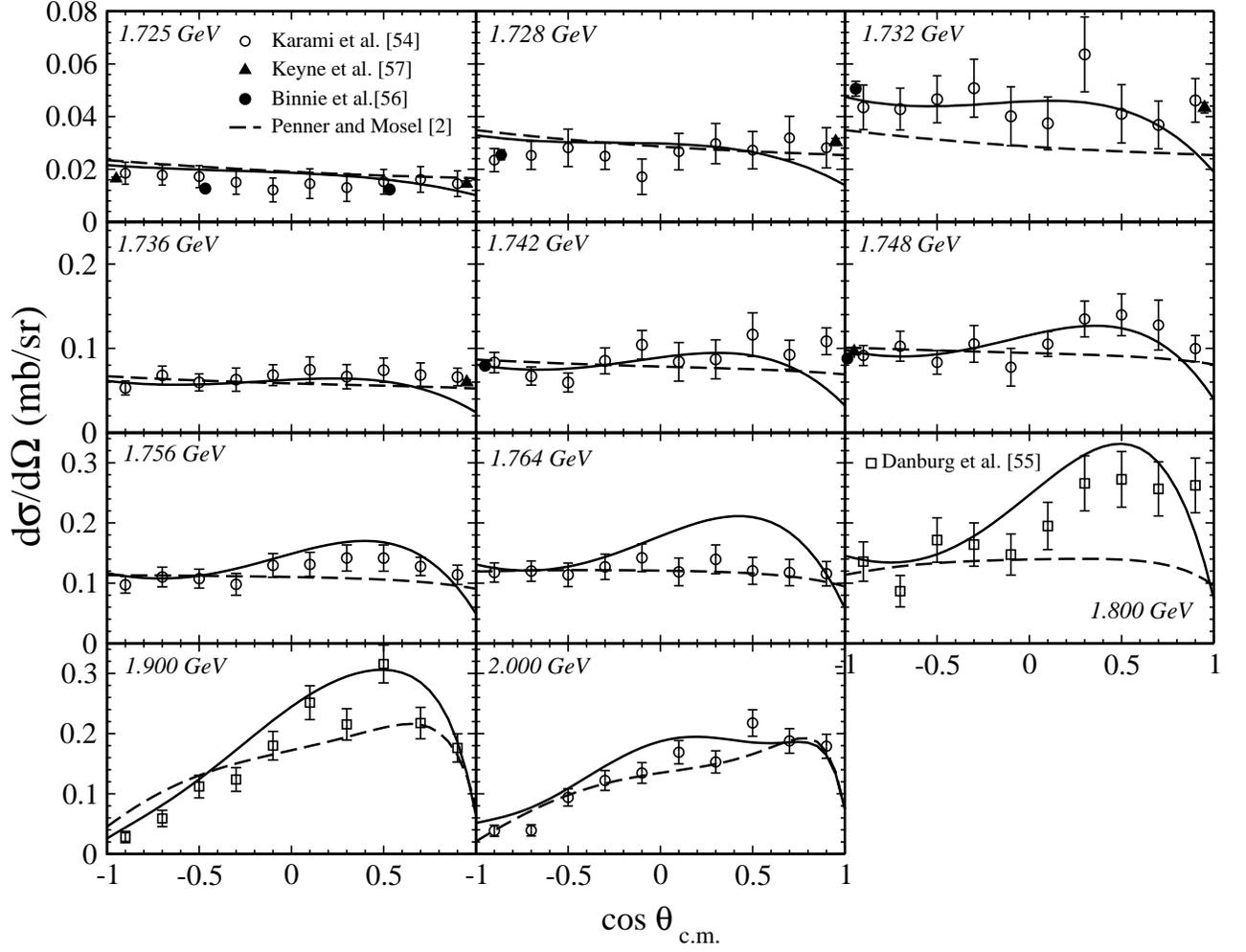}}
       }
       \caption{
Differential cross sections of the  $\pi N \to \omega N$ reaction.
The experimental data are taken from \cite{Karami:1979,Danburg:1971,Binnie:1974,Keyne:1976}.
Our previous best global result from \cite{Penner:2002a} is shown by the  dashed line. 
      \label{omgN_dif}} 
  \end{center}
\end{figure}

\begin{figure}
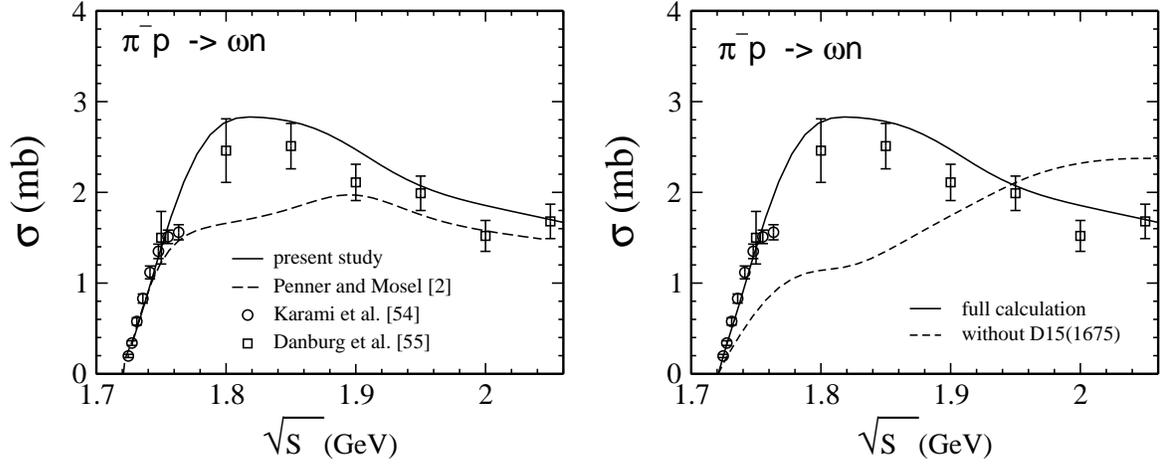

  \begin{center}
    \parbox{16cm}{
      \parbox{78mm}{\includegraphics*[width=78mm]{fig7.eps}}
      \parbox{78mm}{\includegraphics*[width=78mm]{fig8.eps}}
       }
       \caption{
\textit{Left:} The calculated total $\pi N \to \omega N$  cross section  in comparison with 
our previous results from \cite{Penner:2002a}. \textit{Right:} The total cross section
calculated with and without the $D_{15}(1675)$ resonance contributions. 
The experimental data are taken from  \cite{Karami:1979,Danburg:1971}. 
      \label{omgN_tot}} 
  \end{center}
\end{figure}

\begin{figure}
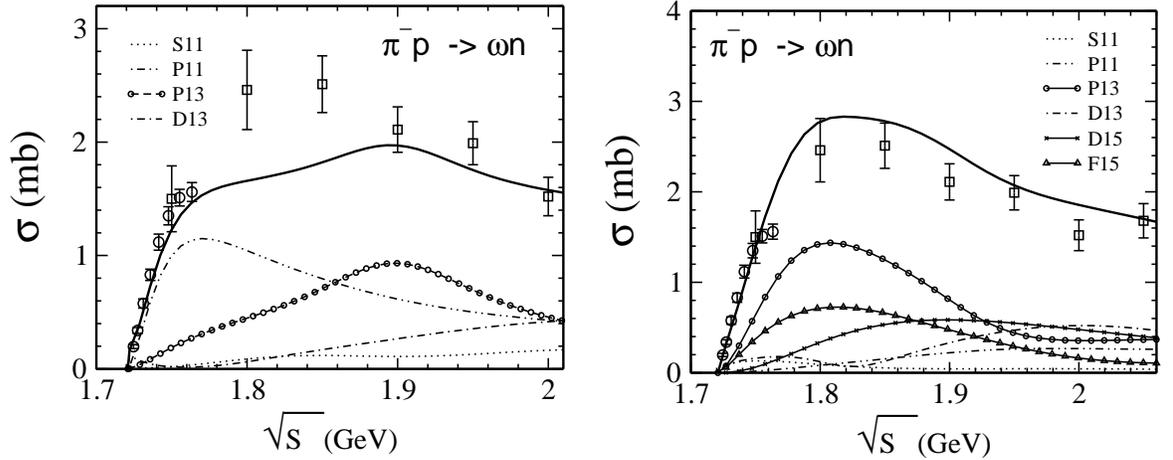

  \begin{center}
    \parbox{16cm}{
      \parbox{78mm}{\includegraphics*[width=78mm]{fig9.eps}}
      \parbox{78mm}{\includegraphics*[width=78mm]{fig10.eps}}
       }
       \caption{
\textit{Left:} The results of the partial wave decomposition of the total $\pi N \to \omega N$  
cross section  from \cite{Penner:2002a}. \textit{Right:} The partial wave cross 
sections obtained in the present calculations.
      \label{omgN_part}} 
  \end{center}
\end{figure}

\section{Results for $\omega$ meson production}
\label{Results}
Our main interest is  the $\omega$-meson production mechanism in the 
pion-  and photon-induced reactions. As pointed out in 
\cite{Penner:2002a,Penner:2002b}, using only the hadronic data is insufficient to determine 
the reaction mechanism. Therefore, we carry out a new combined study of the $\pi N\to \omega N$ and 
$\gamma N \to \omega N$ reactions in up to 2 GeV energy region. The most significant improvements
in the present calculations is the inclusion of the contributions from  spin-$\ffh$ resonances.
In order to constrain the analysis we include the full set of experimentally available informations
into the energy region up to 2 GeV. We expect that this extended analysis will provide a much deeper
insights into the production mechanism as before. 
As compared to the previous calculations \cite{Penner:2002a,Penner:2002b}, 
the additional constraints from the spin density  matrix elements  of the final $\omega$ meson
measured by SAPHIR are also taken into account.

\subsection{$\pi N \to \omega N$}

All experimental data on  the $\omega$-meson production in the $\pi N$ scattering have been 
measured before 1980 and therefore have rather poor statistics. In total, there  are 115 data
points which includes differential and total cross sections data.  
The inclusion of spin-$\ffh$ resonance contributions strongly changes the relative 
resonance contributions to the $\omega N$ final states in the present calculations, see 
Fig. \refe{omgN_dif}. In contrast to the findings in \cite{Penner:2002a}, 
the main contributions close to the threshold  come from the 
$P_{13}$ and $D_{15}$ partial waves. The resonance part of the production amplitude is
dominated by the $D_{15}(1675)$ state. 
The result of our calculations without the $D_{15}(1675)$ 
contribution is shown in Fig. \refe{omgN_tot}.  At the threshold,  the reaction mechanism
is influenced by the $S$-wave contributions leading to a rather flat
angular distribution (see Fig. \refe{omgN_dif}). The major difference between the 
present calculations and  the results from \cite{Penner:2002a} is seen at $\sqrt{s}=$1.80 GeV 
where only the data of Danburg et al. \cite{Danburg:1971} are available. This experimental data shows 
an increase of the differential cross section at forward angles. In our previous calculations
the reaction in this kinematical region is dominated by the $P_{11}$ wave 
contributions resulting  in a weakly angle dependent differential cross sections. The inclusion
of spin-$\ffh$ resonances shifts this strength to the $P_{13}$ and $D_{15}$ partial waves 
and the cross section at $\sqrt{s}=$1.80 GeV follows the Danburg data.

The partial wave decomposition of the $\pi N \to \omega N$ reaction is shown in 
Fig. \refe{omgN_part} in comparison with our previous results \cite{Penner:2002a,Penner:2002b}
where the contributions from the spin-$\ffh$ resonances have been neglected.     
Despite of the significant differences in the production mechanisms, we find 
$\chi^2_{\pi\omega}\simeq=$1.25 in both calculations. 
Thus, the distinction between various results is difficult due to the lack of the 
hadronic data. However, the results in other channels may be used to constrain the reaction 
mechanism. In our previous global fit \cite{Penner:2002a,Penner:2002b}, large 
contributions from the $P_{11}(1710)$ and $P_{13}(1900)$ resonances to this reaction 
have been found. The mass of   $P_{11}(1710)$ ($M$=1752 MeV) has been found to be above the $\omega N$  
threshold  $M$=1.752 so that this resonance dominated the production cross
section from the threshold up to 1.8 GeV. However, such a strong contributions to the 
$\omega N$ channel lead to the excess structure in the real and imaginary parts of the 
$\pi N$  partial wave amplitude $P_{11}$ around 1.73 GeV which is not visible 
by the SAID analysis \cite{Arndt:2003}, see  Section \ref{piN}.  

We also find strong contributions from the $P_{13}$ partial wave to the $\pi N \to \omega N$ 
reaction what has been also reported in  \cite{Penner:2002a}, see Fig. \refe{omgN_part}. 
But in contrast to \cite{Penner:2002a}, the strength in this partial  wave is 
shifted to the lower energies and becomes more pronounced near to the reaction threshold. 
The peaking behaviour in the $P_{13}$ partial cross section is due to the interference pattern
between $P_{13}$ resonances and  background contributions to the $\omega N$ channel. 
Since the major contributions to the $\pi N\to \omega N$ reaction come from the $P_{13}$ and 
$D_{15}$  waves, it is interesting to look at the $\pi N$ inelasticity for these partial 
waves.  The calculated $\sigma_{\pi N}^{inel}$ inelasticity in these waves is found 
to be in line with the  SAID data (see Fig. \refe{2piN_I12}). While the $P_{13}$ inelastic data 
are also well described, it would be also desirable to check the obtained results by including 
contributions from other inelastic ($3\pi N$) channels (see Section \refe{piN}).

\begin{figure}
  \begin{center}
    \parbox{17cm}{
      \parbox{17cm}{\includegraphics*[width=15cm]{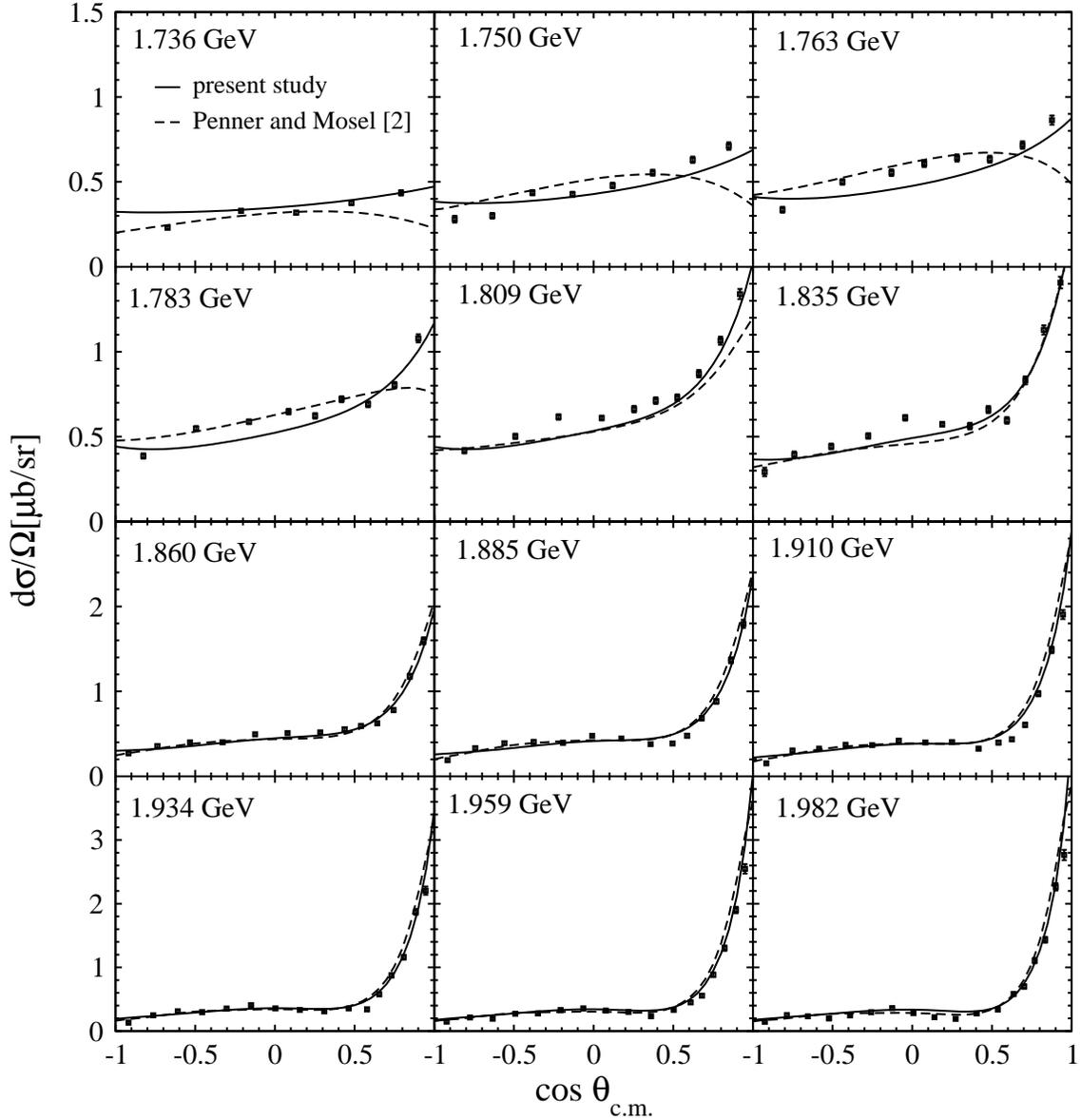}}
       }
       \caption{
$\gamma N \to \omega N$ differential cross sections  in comparison with the 
SAPHIR data \cite{Barth:2003} and our previous results from \cite{Penner:2002b}.
      \label{omgN_photo_dif}} 
  \end{center}
\end{figure}

\begin{figure}
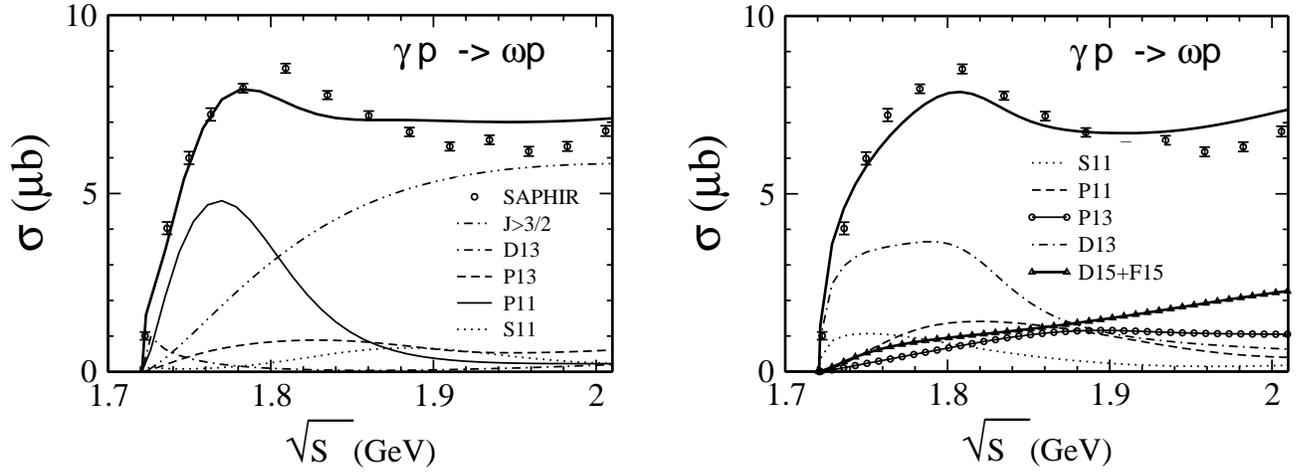

  \begin{center}
    \parbox{18cm}{
      \parbox{8.9cm}{\includegraphics*[width=8.5cm]{fig12.eps}}
      \parbox{8.9cm}{\includegraphics*[width=8.5cm]{fig13.eps}}
       }
       \caption{
\textit{Left:} Total and partial wave cross sections from \cite{Penner:2002b}.
\textit{Right:} Total and partial wave cross sections calculated in the present study.
      \label{omgN_photo}} 
  \end{center}
\end{figure}

\begin{figure}
  \begin{center}
    \parbox{18cm}{
      \parbox{15.cm}{\includegraphics*[width=14.5cm]{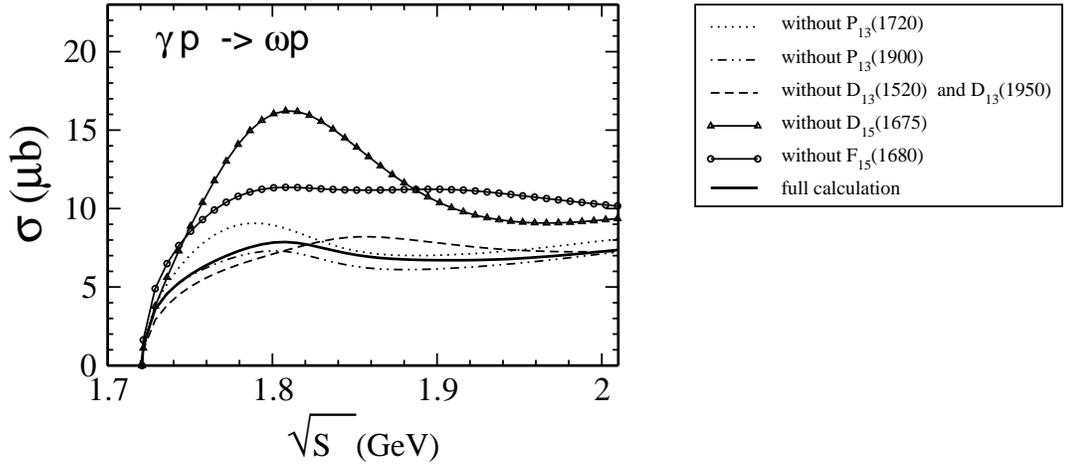}}
       }
       \caption{
Role of the individual resonance contributions in the  $\omega$ meson photoproduction.
      \label{omgN_photo_res}} 
  \end{center}
\end{figure}

\subsection{$\gamma N \to \omega N$}
The differential $\omega$ meson photoproduction  cross sections 
are presented in Fig. \refe{omgN_photo_dif} in comparison with our previous results from 
\cite{Penner:2002a,Penner:2002b}.  In the pesent calculations we obtain $\chi_{\gamma\omega}^2$=4.5
which is significantly better than   our previous result  ($\chi_{\gamma\omega}^2$=6.25). 
This improvement strongly supports the extended treatment applied in this work. 
All studies of this reaction agree on the importance of the  $\pi^0$ exchange reported 
first by Friman and Soyeur \cite{Friman:1995}.
These contributions lead to the peaking behaviour of the calculated differential cross 
sections at forward angles which also becomes visible  in the  SAPHIR measurements \cite{Barth:2003} 
above 1.783 GeV, see Fig. \refe{omgN_photo_dif}. More detailed information of the production mechanism 
is obtained from observables measuring the spin degree of freedom of the $\omega$ meson. 
For the $t$-channel spinless meson exchange ({\it c}-diagram in Fig. \refe{diag}) the 
spin density matrix $\rho_{rr'}$ of the final $\omega$ mesons can be easily calculated, see
Appendix \ref{pi0-exchange}.
In the Gottfried-Jackson frame, where the initial photon and exchange particle are in their 
rest frame, and  $z$-axis is in the direction of the incoming photon momentum, 
the calculation gives $\rho_{00}^{GJ}=0$.
On the other hand, the experimental value of  $\rho_{00}^{GJ}$ for forward directions,
where the $\pi^0$ exchange dominates, was  measured by  SAPHIR and  found to be in the range
of $\rho_{00}^{GJ}=0.2\cdots 0.3$. Thus, the nonzero matrix element testifies that even in this 
kinematical region other mechanisms (rescattering effects, interference with resonances) 
must be important. 

\begin{figure}
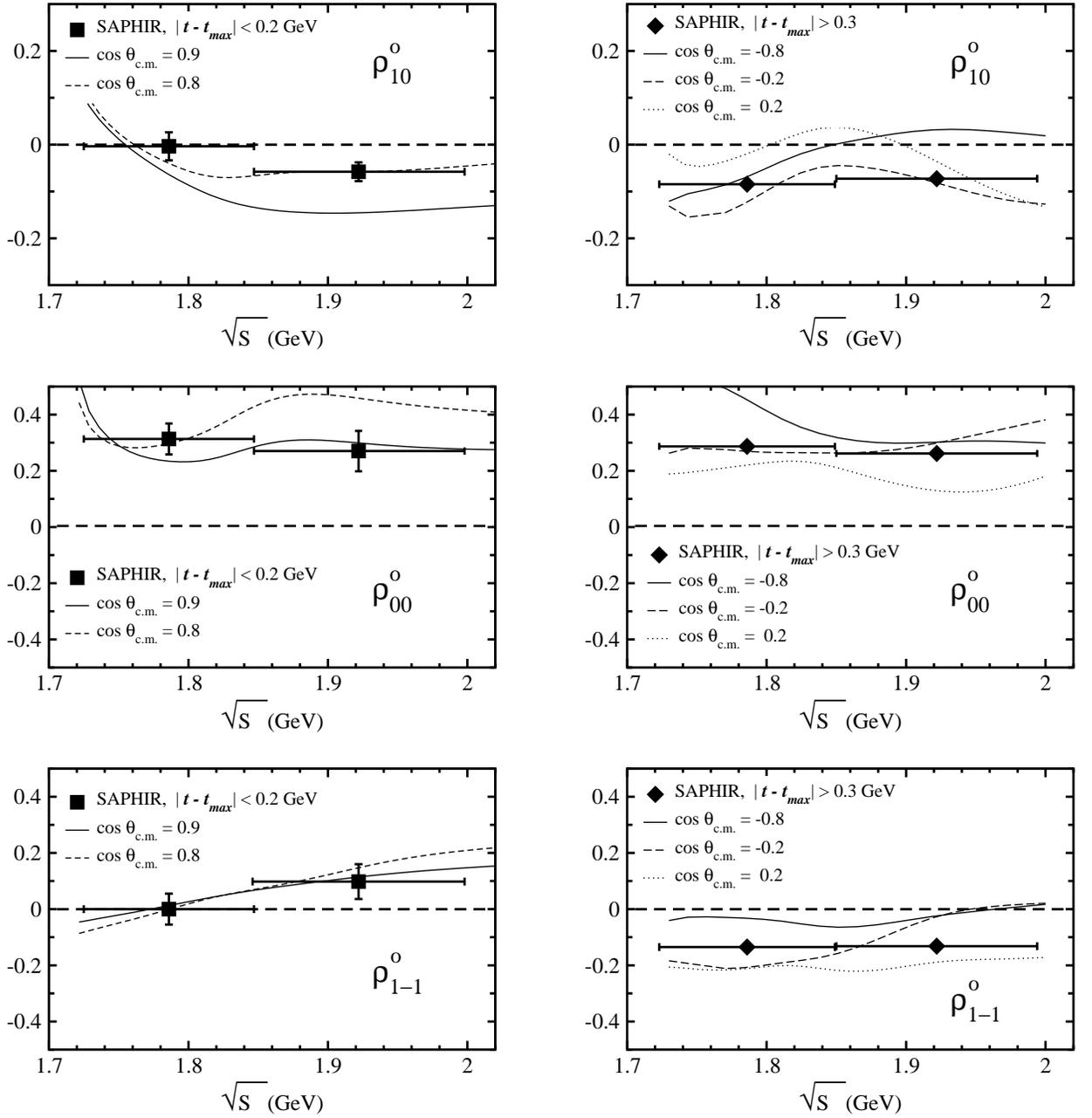

  \begin{center}
    \parbox{19cm}{
      \parbox{8.9cm}{\includegraphics*[width=7.5cm]{fig15.eps}}
      \parbox{8.5cm}{\includegraphics*[width=7.5cm]{fig16.eps}}\\
      \parbox{8.9cm}{\includegraphics*[width=7.5cm]{fig17.eps}}
      \parbox{8.5cm}{\includegraphics*[width=7.5cm]{fig18.eps}}\\
      \parbox{8.9cm}{\includegraphics*[width=7.5cm]{fig19.eps}}
      \parbox{8.5cm}{\includegraphics*[width=7.5cm]{fig20.eps}}\\
       }
       \caption{
Spin density matrix elements in the helicity frame in comparison with  the SAPHIR measurements 
\cite{Barth:2003}.      
\label{omgN_spin}}
  \end{center}
\end{figure}

There is a visible distinction between  results from \cite{Penner:2002b} and present study at 
energies close to $\omega N$ threshold, see Fig. \refe{omgN_photo_dif}. Above 1.835 GeV both 
calculations give  similar results and almost coincide at higher energies. However, the 
differences in the reaction mechanisms can be seen in the partial wave decomposition shown
in Fig. \refe{omgN_photo}.  We find less contributions from the $P_{11}$ partial wave 
as compared to  \cite{Penner:2002b}. However, this channel is still 
important  and gives sizeable  strength near 1.8 GeV. 
In contrast  to \cite{Penner:2002a,Penner:2002b}, the main contribution at the reaction 
threshold comes from  the $D_{13}$ partial wave which also leads to the change
in the differential cross section behaviour, see Fig. \refe{omgN_photo_dif}. Since  
the spin-$\ffh$ resonances were not included in  our previous analysis and at higher energies 
the cross section was entirely dominated by the background contributions from the partial waves with 
spin $J> \fth$, see Fig. \refe{omgN_photo}, left. 

\begin{figure}
  \begin{center}
    \parbox{19cm}{
      \parbox{8.5cm}{\includegraphics*[width=7.1cm]{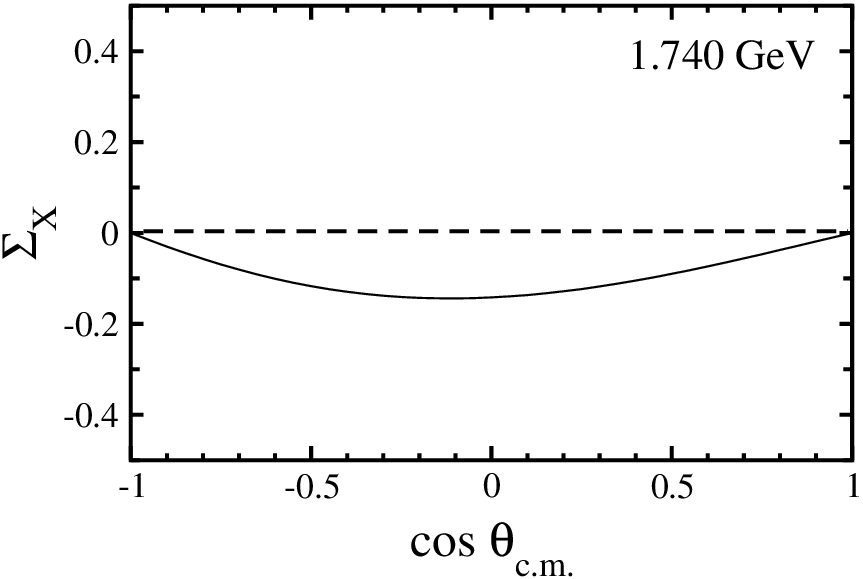}}
      \parbox{8.5cm}{\includegraphics*[width=7.1cm]{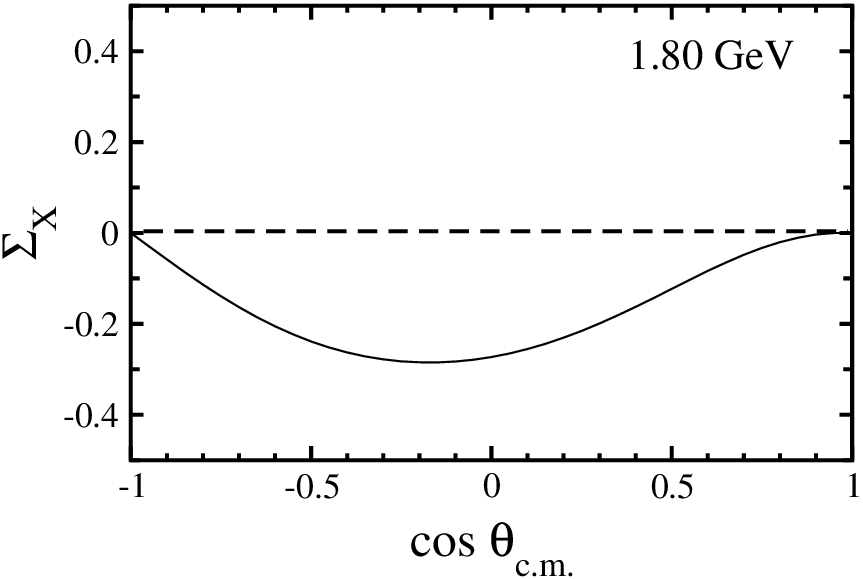}}\\
      \parbox{8.5cm}{\includegraphics*[width=7.1cm]{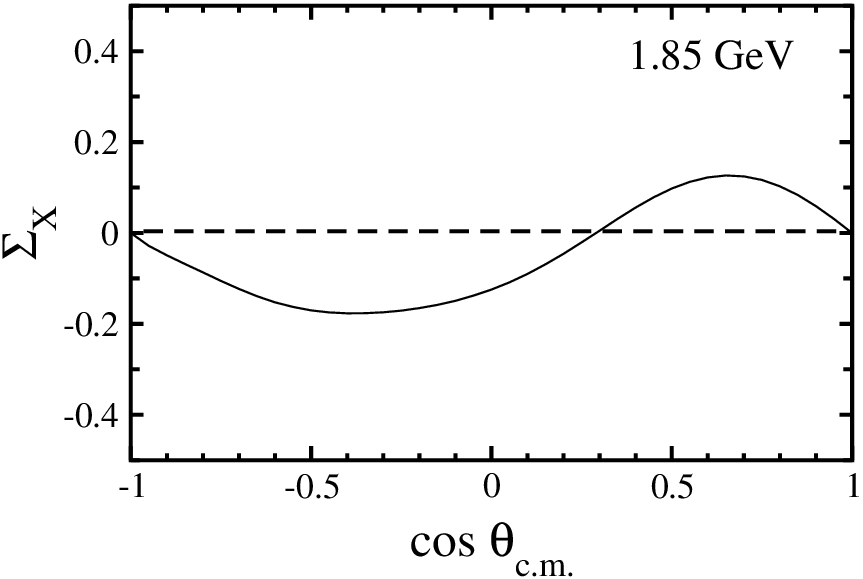}}
      \parbox{8.5cm}{\includegraphics*[width=7.1cm]{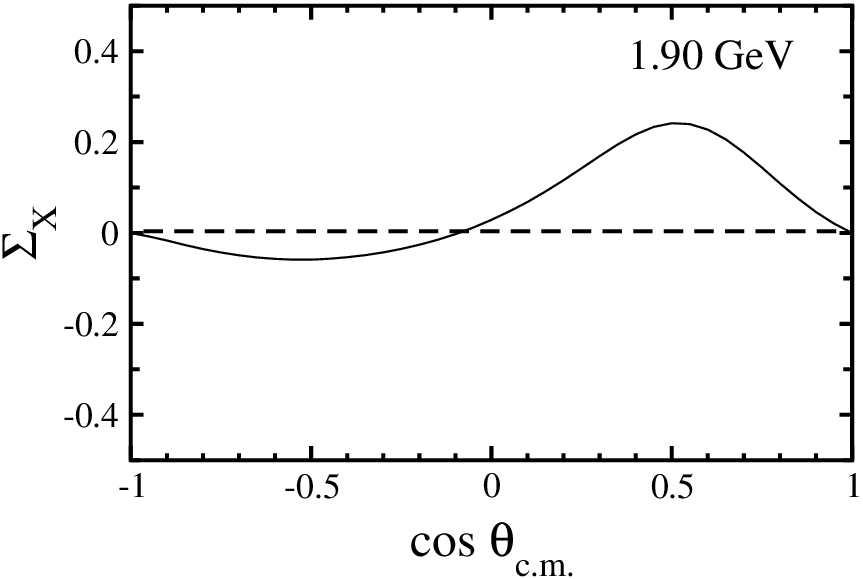}}\\
       }
       \caption{
Photon beam asymmetry $\Sigma_X$ at fixed energies as a function of the $\omega$ production angle.
\label{omgN_asimm}}
  \end{center}
\end{figure}
\begin{figure}
  \begin{center}
    \parbox{19cm}{
      \parbox{8.5cm}{\includegraphics*[width=7.1cm]{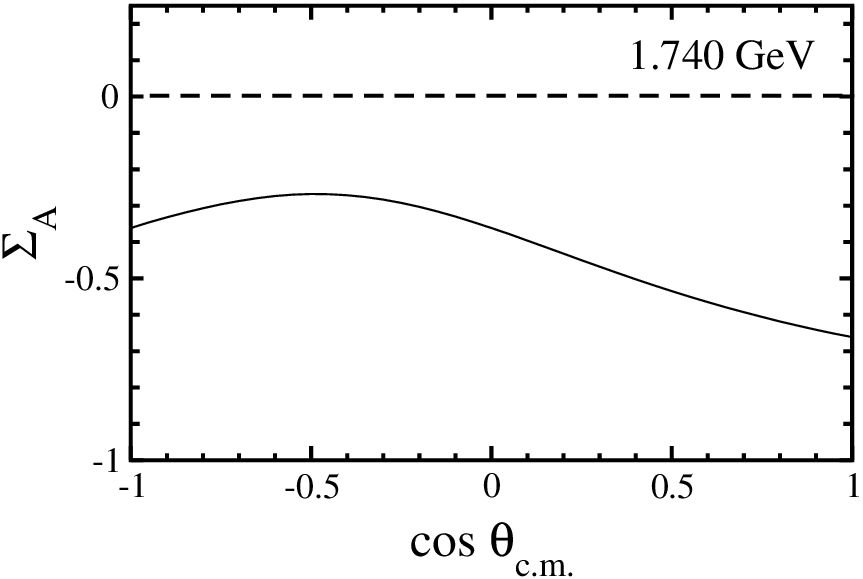}}
      \parbox{8.5cm}{\includegraphics*[width=7.1cm]{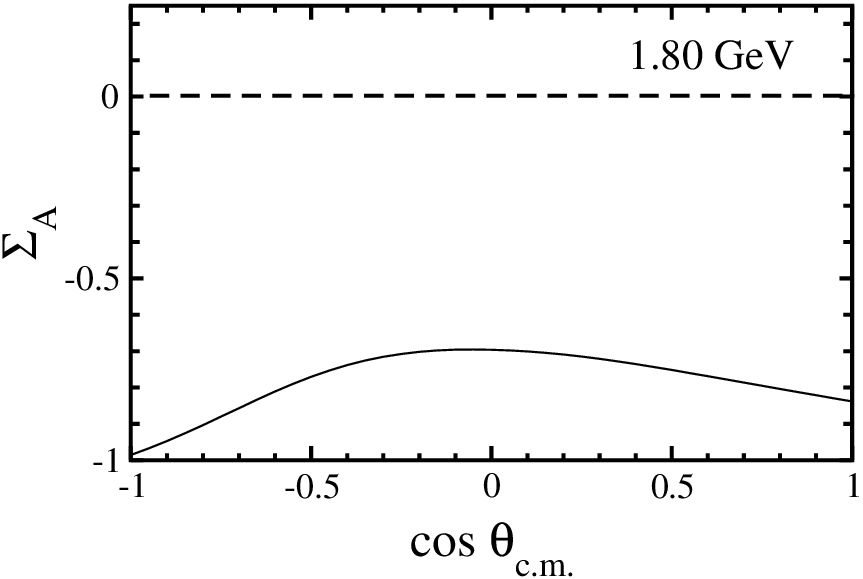}}\\
      \parbox{8.5cm}{\includegraphics*[width=7.1cm]{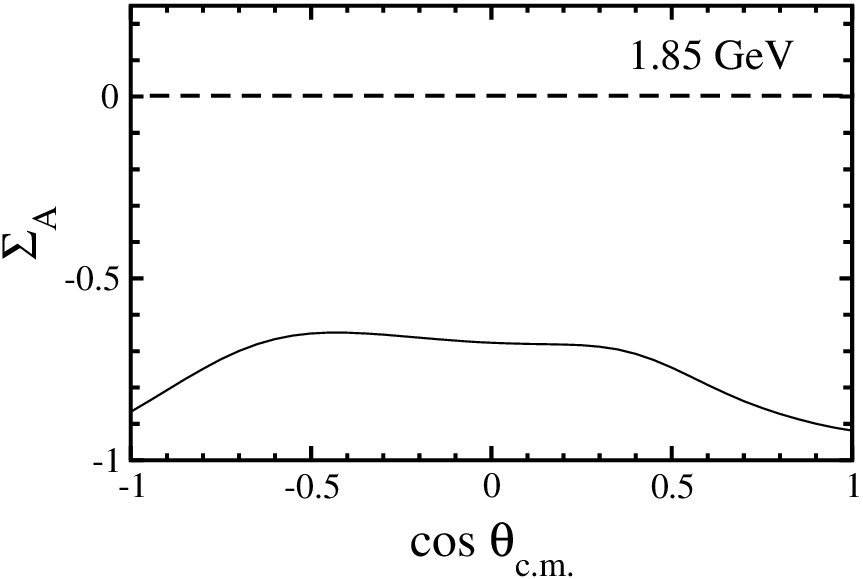}}
      \parbox{8.5cm}{\includegraphics*[width=7.1cm]{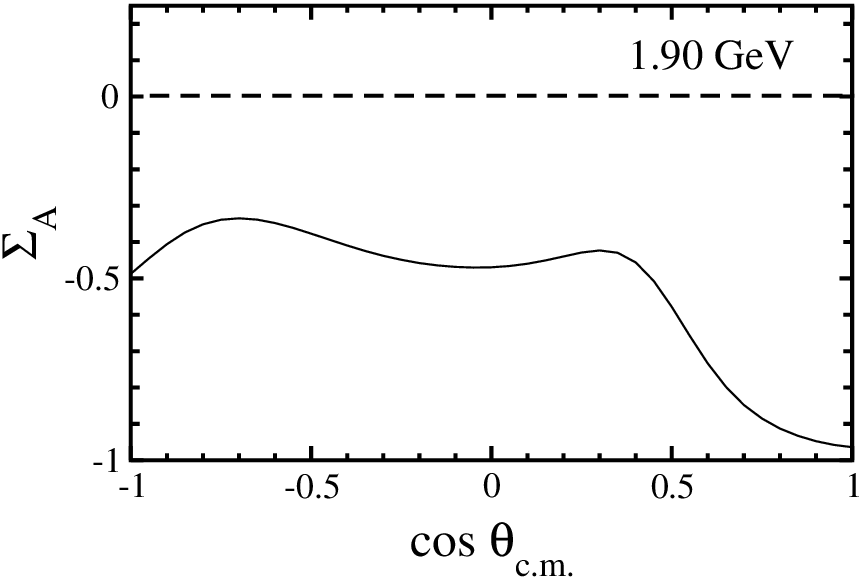}}\\
       }
       \caption{
Photon beam asymmetry $\Sigma_A$ at fixed energies as a function of the $\omega$ production angle.      
\label{omgN_asimmA}}
  \end{center}
\end{figure}

\begin{figure}
  \begin{center}
    \parbox{19cm}{
      \parbox{8.5cm}{\includegraphics*[width=7.1cm]{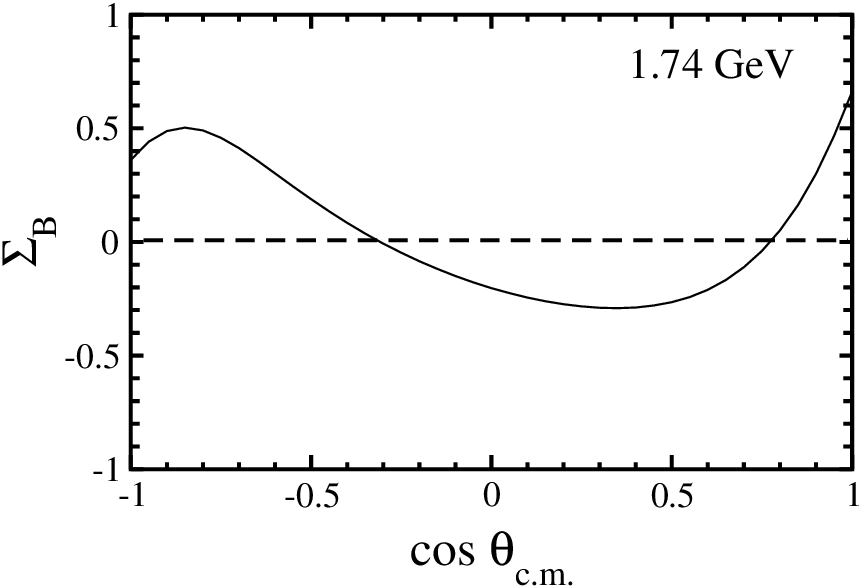}}
      \parbox{8.5cm}{\includegraphics*[width=7.1cm]{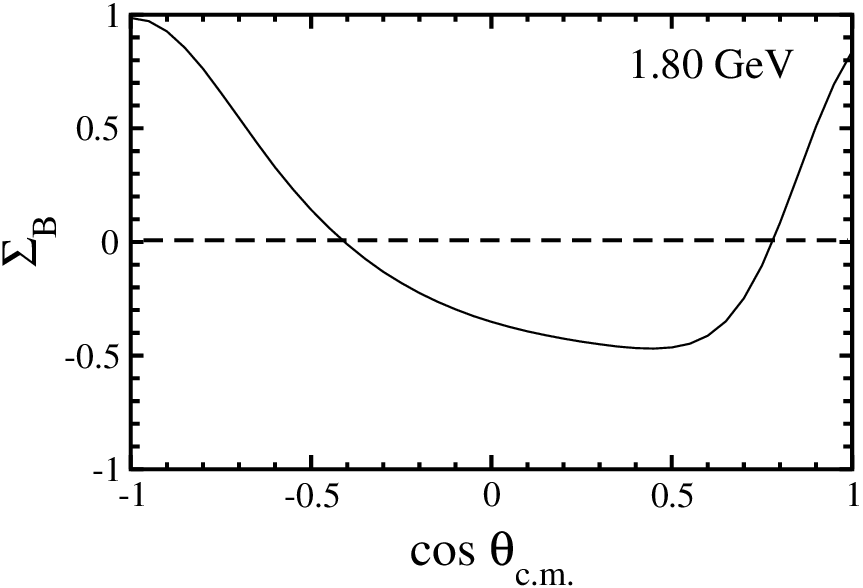}}\\
      \parbox{8.5cm}{\includegraphics*[width=7.1cm]{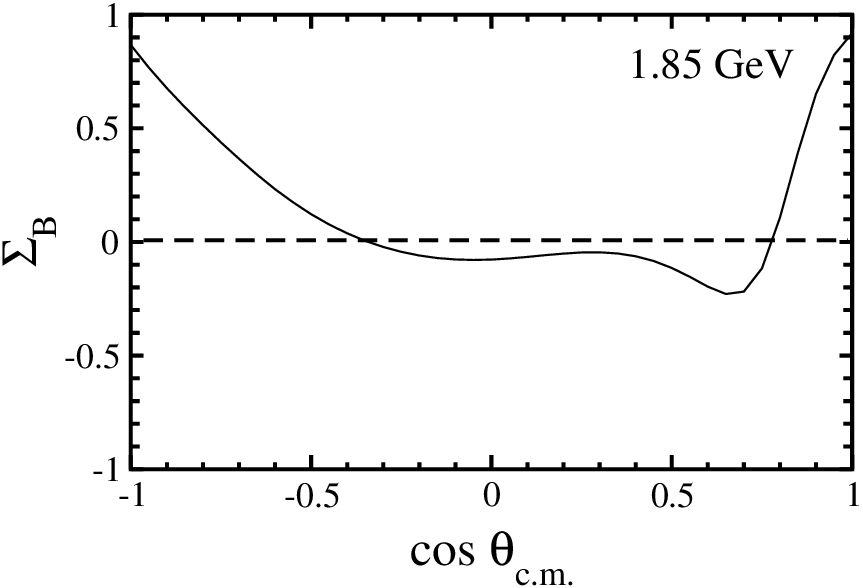}}
      \parbox{8.5cm}{\includegraphics*[width=7.1cm]{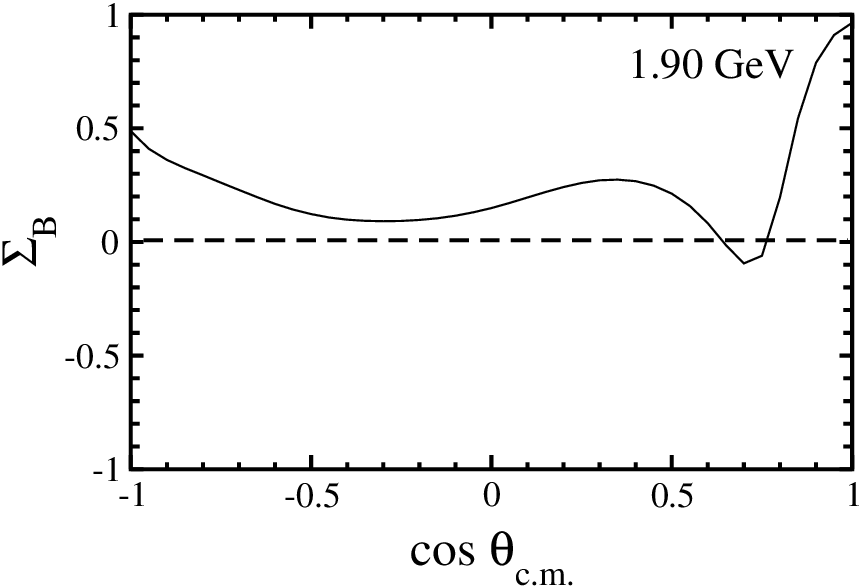}}\\
       }
       \caption{
Photon beam asymmetry $\Sigma_B$ at fixed energies as a function of the $\omega$ production angle.      
\label{omgN_asimmB}}
  \end{center}
\end{figure}

The largest contributions to the $\omega$ meson photoproduction come from the $\pi^0$
exchange and  the subthreshold spin-{$\ffh$} resonances: $D_{15}(1675)$ and $F_{15}(1680)$.
Since the $\pi^0$ exchange  above 1.8 GeV strongly influences the $\gamma N \to \omega N$ 
reaction a consistent identification of individual resonance contributions from the only 
partial wave decomposition shown in Fig. \refe{omgN_photo}, right, is difficult.  
The  $P_{13}(1900)$, and $F_{15}(2000)$, and $D_{13}(1950)$ states which lie above the reaction
threshold hardly influence the reaction due to their small couplings to $\omega N$, 
see Table \ref{tab12}. 
Despite of the small relative contribution from the $D_{15}$ and $F_{15}$ waves to the $\omega$ 
photoproduction the cross sections are strongly affected by spin-$\ffh$ states 
because of the destructive interference  
pattern  between  the $\pi^0$ exchange and  these resonance contributions, 
see Fig. \refe{omgN_photo_res}. 
The $D_{13}(1950)$ state has a large branching ratio into 
the $\omega N$ final state, see Table \ref{tab12}. However, the contributions from this resonance
to the photoproduction are moderate due to its large mass and total width.

While  $F_{15}(1680)$  plays only a minor role in the
$\pi N \to \omega N$ reaction the contribution from this state becomes more pronounced in the
$\omega$ meson photoproduction because of its large $A^p_{\fth}$ helicity amplitude,
as seen in Table \ref{helict}.  
The importance of the  $F_{15}(1680)$ resonance to the $\omega$ meson photoproduction was also  found 
by Titov and Lee \cite{Titov:2002} and in the model of Zhao \cite{Zhao:2000}. However,
in contrast to \cite{Titov:2002} where  also a large effect from  
$D_{13}(1520)$ was observed we do not find any visible contribution from this state.
In fact, the strong contribution to the $D_{13}$ partial  wave seen
in the right panel of Fig. \refe{omgN_photo}, resembling a resonance structure,
comes  indeed  from non-resonant $\pi^0$ exchange.

It is  interesting to note, that both resent study \cite{Titov:2002,Zhao:2000}  
find no significant effect from $D_{15}(1675)$ state in the $\omega$ meson photoproduction. 
Thus, in the quark model of Zhao \cite{Zhao:2000} the contributions from this state is
strictly suppressed by the Moorhouse selection rule \cite{Moorhouse:1966}. While 
Titov and Lee \cite{Titov:2002} account for the $D_{15}(1675)\to \omega N$ contributions
the corresponding $\omega NN^*$ coupling is determined from the VDM assumptions. Since
the electromagnetic helicity amplitudes of this resonance are relatively small, 
see Table \ref{helict}, the resulting  $\omega NN^*$ coupling also has only marginal effect
in this approach.

The  $\rho_{rr'}$  elements extracted from the SAPHIR  data \cite{Barth:2003} 
are an outcome of the averages over  
rather wide energy and angle regions, see Fig. \refe{omgN_spin}. 
Therefore, the original error bars have been decreased by 
factor 3 to put an additional weight to these data. The inclusion of  measured $\rho_{rr'}$
into the calculations  provides a strong additional constraint on the relative partial wave 
contributions and finally on the resonance couplings. 
The spin density matrix elements calculated at fixed angles in the helicity frame are 
presented in Fig. \refe{omgN_spin}. A good description of the spin density matrix is possible in 
a wide energy region.  Since the $\rho_{rr'}$ data put strong constraints on the $\gamma p \to \omega p$
reaction mechanism  there is an urgent need for  precise measurements of the spin density matrix 
in more narrow energy bins to pin down the reaction picture. 

The calculated photon beam asymmetry $\Sigma_X$ (see Appendix \ref{asymm}) is shown in Fig. \refe{omgN_asimm}. 
At energies close to the threshold our calculations predict a negative values of $\Sigma_X$. In the energy
region between $1.72\div 1.8$ GeV the asymmetry has an almost symmetric behaviour. By increasing the c.m. 
energy the $\pi^0$ exchange becomes dominant at forward angles  leading to a change of the sign
at $\Sigma_X$ above 1.85 GeV. Such a  behaviour is especially interesting since it tests an interference 
pattern between the resonance and background  parts of the transition amplitude. 
The role of this interference becomes  more pronounced in the $\Sigma_A$ and $\Sigma_B$ asymmetries,
see Figs. (\ref{omgN_asimmA},\ref{omgN_asimmB}).  For the pure $\pi^0$ exchange production mechanism 
the calculations give   $\Sigma_A$=-1 and $\Sigma_B$=+1, see Appendix \ref{pi0-exchange}.  Therefore
the deviation from these values testifies about magnitude of the interference between the
$\pi^0$ exchange  and other production mechanisms. The experimental measurements of these observables 
provide a good test for the presented model.

\begin{figure}
  \begin{center}
    \parbox{18cm}{
      \parbox{8.9cm}{\includegraphics*[width=8.5cm]{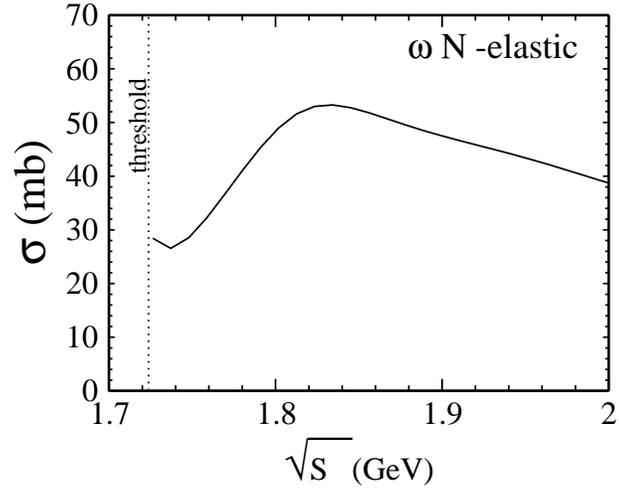}}
       }
       \caption{
Elastic $\omega N$ scattering calculated in the present study.
      \label{omgN_elast}} 
  \end{center}
\end{figure}

\section{$\omega N$ elastic scattering}
In the present calculations  this process is completely dominated by the nucleon resonance 
contributions. The effect of the nucleon Born term is only marginal.
The calculated $\omega N$ elastic total cross section is shown in Fig. \refe{omgN_elast}. 
The extracted scattering lengths and effective ranges (as defined in Appendix \ref{scattering_lengths}) 
are
\bea
\ba{lcrclcr}
\bar a   &=& -0.026  + \mi 0.28\;, & & 
\bar r   &=& 6.02 + \mi 0.062\;. \\
\ea
\label{ScattLength}
\eea
The result for $Im \bar a$=0.28 fm is consistent with values obtained by Lutz et al. 
\cite{Lutz:2001} $\bar a = -0.44  + \mi 0.20$ fm
and Klingl and Weise \cite{Klingl:1998} $a = 1.6  + \mi 0.30$ fm. Note, however that  calculated scattering
lengths \refe{ScattLength} should be taken with care since the present calculations are not concentrated on 
the description of the near to threshold region but consider a rather wide energy range.
Since the resonance $\omega N$ couplings are constrained by the $\pi N\to \omega N$ and $\gamma N \to \omega N$ data,
the extracted scattering lengths might suffer from the  lack of experimental information at the $\omega N$ threshold.

\section{Summary}
\label{summary}
In the present study we perform a new analysis of the $\omega$ meson production in $\pi N$ and
$\gamma N$ reactions within a unitary effective Lagrangian coupled-channel formalism.
We have investigated contributions to the $\omega N$ final state from all spin-$\foh$,-$\fth$, and spin-$\ffh$
resonances with masses below 2 GeV. To fix the resonance couplings a coupled-channel calculation has 
been carried out for the final states $(\gamma/\pi) N \to \gamma N$, $\pi N$, $2\pi N$, $\eta N$, 
and $\omega N$ where free parameters of the model are constrained by the all available experimental
reaction data for energies from the pion threshold and up to 2 GeV. 
The extracted resonance 
couplings to $\gamma N$, $\pi N$, $2\pi N$, and $\eta N$ are found in a good agreement with the
results from other analyses and PDG.

Because of the inclusion of the spin-$\ffh$ resonance contributions we obtain 
a significantly better 
description of the $\omega$ photoproduction data as compared to our previous calculations.
The experimental data on the spin density matrix elements  $\rho_{00}$, $\rho_{10}$, and 
$\rho_{1-1}$ measured by SAPHIR,  give important constraints on the $\omega$ meson production 
mechanism.  
We find a strong contribution from the $D_{15}(1675)$ resonance to the 
$\omega N$ final state 
in the pion- and photon-induced reactions. While the $F_{15}(1680)$ state hardly influences the  
$\pi N \to \omega N$ process the contribution from this resonance to the $\omega$ meson 
photoproduction turns out to be significant due to its large $A_{\fth}^p$ helicity amplitude.
A strong contribution to the $\omega$ photoproduction  comes from the $D_{13}$ partial wave 
which is  dominated by the $\pi^0$ exchange. The effect from the $D_{13}(1520)$ and $D_{13}(1950)$ 
states in this reaction is of minor importance.

Apart from the $D_{15}(1675)$ and $F_{15}(1680)$ resonance contributions the $\omega$ meson
photoproduction is strongly dominated by the $\pi^0$ exchange mechanism which has been also
found in  the previous findings. 
We conclude that for the correct description of the experimental data on the $\omega$ meson 
production the contribution from the nucleson resonances should be taken into account.
However, due to the strong interference pattern between 
resonances and  the $\pi^0$ exchange the separation of the individual resonance contributions
is difficult in this reaction.
Hence, a search for 'hidden' resonances  with the help of this channel 
becomes questionable. 
We predict a negative sign of the photon beam asymmetry in the $\omega$ photoproduction  at
energies close to the threshold. Above the 1.85 GeV the $\Sigma_X$ asymmetry changes its behaviour and 
becomes positive at the forward angle directions.
Since polarization observables, such as $\rho_{rr'}$ and $\Sigma_X$,$\Sigma_A$, and $\Sigma_B$ 
are very sensitive to different reaction mechanisms the  more precise measurements  of these 
quantities  are urgently needed to distinguish between various model predictions. 
Our predictions for these observables can easily be checked at 
GRAAL, CLAS, and CB-ELSA.  
 
\begin{appendix}

\section{Couplings, decay widths, and helicity amplitudes}
\label{applagr}
All interaction Lagrangians for spin-$\foh$,$\fth$ resonances and Born terms can be found
in \cite{Penner:2002a,Penner:2002b,Penner:PhD}.
Here,  we list only  those couplings which
are extensions of our previous calculations keeping, however, the same notations.  
The Lagranginas   given below contain
an isospin part, which is also  discussed \cite{Penner:2002a,Penner:2002b,Penner:PhD}
together with the isospin and  partial wave decomposition.

\subsection{Tensor meson coupling}
\label{tensor_couplings}
The coupling of  the tensor $f_{2}(1270)$ meson to the nucleon field is described by  
\bea
\mcl_{NNf_2}  = - \bar u_{N'} (p') \bigg[ 
-\mi\frac{g_{NNf_2}}{m_N}(\gamma_\mu
\bar{\partial}_\nu^{(N'N)}+ \gamma_\nu \bar{\partial}_\mu^{(N'N)})f^{\mu \nu}_2
+\frac{h_{NNf_2}}{m_N^2}\bar{\partial}_\mu^{(N'N)}\bar{\partial}\nu^{(N'N)}f^{\mu \nu}_2
\bigg ] u_N (p)
\label{lagback} 
\eea
with the asymptotic nucleons $N,N'=N$.
The notation $\bar{\partial}_\mu^{(N'N)}$
in the Eq. \refe{lagback}, related to the  nucleon-(tensor)meson $f^{\mu \nu}$ coupling means
$\bar{\partial}_\mu^{(N'N)}= \partial_\mu^{(N)}-\partial_\mu^{(N')}$ where $\partial_\mu^{(N)}$
is a derivative over the nucleon field. Such a definition leads to the same transition  amplitudes
as defined in \cite{Pilkuhn:1973,Goldberg:1968}.

The decay $f_2(1270)\to \pi\pi$ is described by the Largrangian  
\bea
\mcl_{f_2\pi\pi} = -\frac{2g_{f_2\pi\pi}}{m^2_{f_2}}
\partial_\mu \pi \partial_\nu \pi f_2^{\mu\nu},
\eea
which is similar to the form  used in  \cite{Pilkuhn:1973,Oh:2003}.
This leads to the vertex function for the $f_2\to\pi\pi$ decay: 
\bea
V_{f_2\pi\pi}= -\frac{g_{f_2\pi\pi}}{m_f^2}(p_1-p_2)_\mu (p_1-p_2)_\nu \xi ^{\mu \nu}.
\eea
$p_1$ and  $p_2$ are outgoing pion momenta and $\varepsilon ^{\mu \nu}$ is a polarization
vector of the tensor meson.
Using the spin-2 projection operator 
(see \cite{Sharp:1963,Weinberg:1964,Bellucci:1994,Toublan:1995}) 
\bea
P_{\mu\nu;\rho\sigma}(p)= 
&-&\frac{1}{3}\left(-g_{\mu\nu}+\frac{p_\mu p_\nu}{m_p^2}      \right) 
              \left(-g_{\rho\sigma}+\frac{p_\rho p_\sigma}{m_p^2}\right) \nonumber \\
&+&\frac{1}{2}\left(-g_{\mu\rho}+\frac{p_\mu p_\rho}{m_p^2}    \right) 
              \left(-g_{\nu\sigma}+\frac{p_\nu p_\sigma}{m_p^2}\right)  \nonumber\\
&+&\frac{1}{2}\left(-g_{\mu\sigma}+\frac{p_\mu p_\sigma}{m_p^2}\right) 
              \left(-g_{\nu\rho}+\frac{p_\nu p_\rho}{m_p^2}    \right). 
\eea
The decay width $f_2(1270)\to \pi\pi$ is then given by 
\bea
\Gamma = \frac{g^2_{f_2\pi\pi}}{80\pi} m_f \left(1-4\frac{m_\pi^2}{m_f^2}\right)^\ffh.
\eea

\subsection{Spin-$\ffh$ Baryon Resonance Interactions}
\label{appres52lagr}

For practical calculations we adopt the spin-$\ffh$ propagator in the form 
\bea
\mcp_\ffh^{\mu\nu,\rho \sigma}(p) = 
\frac{( p\hskip-0.5em /\ +m_p)}{p^2-m_p^2+ \mi\epsilon}P_{\ffh}^{\mu\nu,\rho \sigma}(p),
\eea
with
\bea
P_{\ffh}^{\mu\nu,\rho \sigma}(p)&=&\frac{1}{2}(
 T^{\mu\rho}T^{\nu \sigma}
+T^{\mu \sigma}T^{\nu \rho} )
-\frac{1}{5}2T^{\mu\nu}T^{\rho \sigma}\nonumber\\
&+&\frac{1}{10}(
T^{\mu\lambda} \gamma_\lambda\gamma_\delta T^{\delta \rho}T^{\nu \sigma}
+T^{\nu\lambda} \gamma_\lambda\gamma_\delta T^{\delta \sigma}T^{\mu \rho}
+T^{\mu\lambda} \gamma_\lambda\gamma_\delta T^{\delta \sigma}T^{\nu \rho}
+T^{\nu\lambda} \gamma_\lambda\gamma_\delta T^{\delta \rho}T^{\mu \sigma}),
\label{ConvPr}
\eea
and
\bea
T^{\mu\nu}&=&-{\rm g}^{\mu\nu}+\frac{p^\mu p^\nu}{m_p^2},
\label{ConvPr2}
\eea
which has also  been  used in an analysis of  $K\Lambda$
photoproduction \cite{David:1995}.

\subsubsection{(Pseudo-)Scalar Meson Decay}

The Lagrangian for the positive parity spin-$\ffh$ resonance decay to a final nucleon 
$N$ and a (pseudo-)scalar  
meson $\varphi$ is chosen in the form
\bea
\mcl_{\ffh N\varphi } = \frac{{\rm g}_{R N \varphi }}{m_\pi^2}\bar u_R^{\mu \nu}
\Theta_{\nu\lambda}(a_{RN\varphi}) 
\left( \begin{array}{c} -\mi \gamma_5 \\ 1 \end{array} \right)
  u_N  \partial_\mu \partial^\lambda 
\varphi + h.c.,
\label{Lagran52}
\eea

and for the negative-parity resonances

\bea
\mcl_{\ffh N\varphi } = -\frac{{\rm g}_{R N \varphi }}{m_\pi^2}\bar u_R^{\mu \nu}
\Theta_{\nu\lambda}(a_{RN\varphi}) 
\left( \begin{array}{c} 1 \\ \mi \gamma_5 \end{array} \right)
  u_N  \partial_\mu \partial^\lambda 
\varphi + h.c.,
\eea

where the upper (lower) factor corresponds to  pseudoscalar (scalar) mesons $\varphi$.

The free spin-$\ffh$ Rarita-Schwinger symmetric field  
$u_R^{\mu \nu}$ obeys the Dirac equation and satisfies the conditions 
$\gamma_\mu u_R^{\mu \nu}=\partial_\mu u_R^{\mu \nu} ={\rm g}_{\mu\nu} u_R^{\mu \nu} =0$ 
\cite{Rarita:1941}. The off-shell projector $\Theta_{\mu\nu}(a)$
is 
\bea
\Theta_{\mu\nu}(a) = g_{\mu\nu}- a\gamma_\mu\gamma_\nu,
\label{off_shell_proj}
\eea
where $a$ is related to the commonly used off-shell parameter $z$
by $a = (z + \sfoh )$.

These couplings lead to the decay width
\refe{Lagran52}
are 
\bea
\Gamma_\pm^\ffh = f_I\frac{{\rm g^2_{R N\varphi }}}{30\pi m_\pi^4}k_\varphi^5
\frac{E_N\mp m_N}{\sqrt{s}}.
\label{width}
\eea
The upper sign corresponds to the decay of the resonance into a meson with the  
identical parity and
vice versa. The isospin factor $f_I$ is the same as for spin-$\foh$,$\fth$ resonances (see 
\cite{Penner:2002a}). $k_\varphi$, $E_N$, and $m_N$ are the meson  momentum, 
energy and mass of the final nucleon, respectively.

\subsubsection{Vector Meson Decay}

The coupling  of the spin-$\ffh$ resonances to the   
$\omega N$ final state is chosen  to be 
\bea
\mcl_{\ffh N \omega} =
\bar u_R^{\mu \lambda}
\left( \begin{array}{c} 1 \\ \mi \gamma_5 \end{array} \right)
\left( \frac{{\rm g}_1}{4m_N^2}\gamma^\xi
+\mi\frac{{\rm g}_2}{8m_N^3}\partial^\xi_{N}
+\mi\frac{{\rm g}_3}{8m_N^3}\partial^\xi_{\omega}\right)
(\partial^{\omega}_\xi{\rm g}_{\mu\nu} - 
\partial^{\omega}_\mu{\rm g}_{\xi\nu})u_N \partial_\lambda^\omega  \omega^\nu + h.c.,
\label{Lagr52om}
\eea
where the upper (lower) factor corresponds to  positive (negative) parity resonances 
and $\partial^\mu_N$ ($\partial_\mu^\omega$)  denotes the partial derivative of
the nucleon  and the $\omega$-meson  fields, respectively.  
Note that, in the spin-$\fth$ case,
the couplings are also contracted by an off-shell projector
\refe{off_shell_proj}.
 Similar coupling was also used to describe 
electromagnetic processes \cite{Zetenyi:2001,David:1995,Titov:2002}.
The couplings \refe{Lagr52om} lead to the helicity-decay amplitudes 
\bea
A^{\omega N}_{\fth} &=&
\frac{\sqrt{E_N\pm m_N}}{\sqrt{5m_N}}
\frac{k_\omega}{4m_N^2}
\left (
-{\rm g}_1(m_N\mp m_R)
+{\rm g}_2\frac{(m_R E_N-m_N^2)}{2m_N}
+{\rm g}_3\frac{m_\omega^2}{2m_N} \right ),  \nonumber\\
A^{\omega N}_{\foh} &=&
\frac{\sqrt{E_N\pm m_N}}{\sqrt{10m_N}}
\frac{k_\omega}{4m_N^2}
\left ( 
{\rm g}_1 (m_N\pm (m_R-2E_N)) 
+{\rm g}_2\frac{(m_R E_N -m_N^2)}{2m_N}
+{\rm g}_3\frac{m_\omega^2}{2m_N} \right ),  \nonumber\\
A^{\omega N}_{0}&=&\frac{\sqrt{(E_N\pm m_N)}}{\sqrt{5m_N}}
\frac{k_\omega m_\omega}{4m_N^2} 
\left ( {\rm g}_1
\pm {\rm g}_2\frac{E_N}{2m_N}
\pm {\rm g}_3\frac{(m_R-E_N)}{2m_N} \right ),
\label{helic}
\eea
with upper (lower) signs corresponding to positive (negative) resonance parity. The lower 
indices stand for the helicity $\lambda$ of the final $\omega N$ state
$\lambda=\lambda_V -\lambda_N$ where 
we use  an
abbreviation as follows: $\lambda=$ $0: 0+\foh$, ~$\foh:1-\foh$,
~ $\fth:1+\foh$.

\subsubsection{Radiative decay}
The coupling  of the spin-$\ffh$ resonances to the   
$\gamma N$ final state is chosen  to be 
\bea
\mcl_{\ffh N \gamma} =
e\bar u_R^{\mu \lambda}
\left( \begin{array}{c} 1 \\ \mi \gamma_5 \end{array} \right)
\left( \frac{{\rm g}_1}{4m_N^2}\gamma^\nu
+\mi\frac{{\rm g}_2}{8m_N^3}\partial^\nu_{N}\right)
u_N \partial_\lambda  F_{\nu\mu} + h.c.,
\eea
where the upper (lower) factor corresponds to positive (negative) parity resonances. 
Note that, 
both couplings are also contracted by an off-shell projector
\refe{off_shell_proj}.
Similar coupling was also used to 
describe  electromagnetic processes \cite{Zetenyi:2001,David:1995,Titov:2002}.

The electromagnetic helicity amplitudes, which are
normalized by an additional factor $(2 E_\gamma)^{-\foh}$
\cite{Warns:1989}, are extracted:
\bea
A^{\gamma N}_\foh &=& 
+ \frac{e \xi_R}{8 m_N^2}
\frac{\sqrt{m_R^2 - m_N^2}}{\sqrt{5 m_N}}
\left(\frac{m_R^2-m_N^2}{2m_R}\right)
\left( g_1 \frac{m_N}{m_R} + g_2 \frac{m_R \pm m_N}{4 m_N} \right) 
\; ,
\nonumber \\
A^{\gamma N}_\fth &=& 
+ \frac{e \xi_R}{4 m_N^2}
\frac{\sqrt{m_R^2 - m_N^2}}{\sqrt{10 m_N}}
\left(\frac{m_R^2-m_N^2}{2m_R}\right)
\left( g_1 + g_2 \frac{m_N \pm m_R}{4 m_N} \right),
\eea
for spin-$\ffh$ resonances. The upper (lower) sing corresponds to positive 
(negative) parity resonances. $\xi_R$ denotes the phase at
the $RN\pi$ vertex. The lower indices correspond to the $\gamma N$
helicities and are determined by the $\gamma$ and nucleon helicities:
$\foh$: $\lambda_\gamma - \lambda_N = 1 - \foh = \foh$ and $\fth$: $1
+ \foh = \fth$.

\subsection{off-shell parameters}

The off-shell parameters used at the interaction vertecies are shown in Table \ref{off_shell_param}.
To reduce a number of free parameters of the model we use one overall 
off-shell parameter for the $R^\ffh N\omega$ couplings, so that  
${a_{\omega N}}_3={a_{\omega N}}_2={a_{\omega N}}_1$.
 
\begin{table}[h]
  \begin{center}
    \begin{tabular}
      {l|r|r|r|r|r|r|r|r}
      \hhline{=========}
      $L_{2I,2S}$ &$a_{\gamma_1}$  & $a_{\gamma_2}$ & $a_{\pi N}$ & $a_{\zeta N}$ & $a_{\eta N}$ &  
      ${a_{\omega N}}_1$ & ${a^b_{\omega N}}_2$ & ${a^b_{\omega N}}_3$ \\ 
      \hhline{=========}
$P_{13}(1720)$ &  2.000 &  -1.273 &  -0.650 &   0.581 &   2.000  &   1.404 &  -1.000 &   0.951 \\
\hline        
$P_{13}(1900)$ & -3.480 &  -0.998 &  -2.000 &   0.643 &   1.977  &   2.537 &  -2.000 &   1.483 \\
      \hhline{=========}
$D_{13}(1520)$ &  0.566 &  0.811  &   0.007 &   0.803 &   0.687  &  -1.000 &   2.000 &   2.000 \\
      \hline        
$D_{13}(1950)$ &  1.389 &  1.440  &  -0.238 &   0.069 &  -2.000  &   0.529 &  -1.999 &   0.108 \\
      \hhline{=========}
$D_{15}(1675)$ &  0.882 &  1.000  &   0.313 &   0.198 &  -1.500  &   1.398 &   ---   &   ---   \\
      \hhline{=========}
$F_{15}(1680)$ & 0.408  &  0.955  &   0.179 &   0.006 &   0.387  &  -0.851 &   ---   &   ---   \\
      \hline        
$F_{15}(2000)$ & 1.000  &  0.089  &   1.697 &  -0.426 &   0.999  &  -0.545 &   ---   &   ---   \\
      \hhline{=========}
    \end{tabular}
  \end{center}
  \caption{Off-shell parameters $a$ of the spin-$\fth$,$\ffh$
    resonances. $^b$: ${a_{\omega N}}_3={a_{\omega N}}_2={a_{\omega N}}_1$ for spin-$\ffh$ resonances.
    \label{off_shell_param}}
\end{table}

\section{Observables}
\label{appobs}

\subsection{Spin density matrix}
\label{sdm}

The spin density matrix of the final  $\omega$ mesons produced in the unpolarized $\gamma N\to \omega N$ 
reaction is written as follows:  
 
\bea
\rho^0_{\lambda_{\omega}\lambda_\omega'} = 
\frac{\sum_{\lambda_N,\lambda_{N'},\lambda_\gamma} 
T_{\lambda_{\omega}\lambda_{N'},\lambda_\gamma\lambda_N}
T^*_{\lambda_{\omega'}\lambda_{N'},\lambda_\gamma\lambda_N}}
{\sum_{\lambda_N,\lambda_{N'},\lambda_\gamma\lambda_\omega} 
T_{\lambda_{\omega}\lambda_{N'},\lambda_\gamma\lambda_N}
T^*_{\lambda_{\omega}\lambda_{N'},\lambda_\gamma \lambda_N}}.
\label{rho0} \; 
\eea
$\lambda_{N}$,$\lambda_{N'}$=$\pm\foh$ stand for the helicity of the initial and final nucleon. 
$\lambda_\omega$=$\pm 1$,$0$ and  $\lambda_\gamma$=$\pm 1$ correspond to the $\omega$ meson and 
photon helicity respectively. 
For polarized reactions one can define 
\bea
\rho^1_{\lambda_{\omega}\lambda_\omega'} = 
\frac{\sum_{\lambda_N,\lambda_{N'},\lambda_\gamma} 
T_{\lambda_{\omega}\lambda_{N'}, -\lambda_\gamma\lambda_N}
T^*_{\lambda_{\omega'}\lambda_{N'},\lambda_\gamma\lambda_N}}
{\sum_{\lambda_N,\lambda_{N'},\lambda_\gamma\lambda_\omega} 
T_{\lambda_{\omega}\lambda_{N'},\lambda_\gamma\lambda_N}
T^*_{\lambda_{\omega}\lambda_{N'},\lambda_\gamma \lambda_N}}.
\label{rho1} \; 
\eea

\subsection{Beam asymmetry for the $\omega$ photoproduction}
\label{asymm}

The photon beam asymmetry for the meson photoproduction reactions is defined as
 
\bea
\Sigma_X = \frac{d\sigma_\perp - d\sigma_\parallel}{d\sigma_\perp + d\sigma_\parallel}
\label{Basymm} \; ,
\eea
where  $d\sigma_\parallel$ ($d\sigma_\perp$) is a differential cross section  of the
$\gamma p\to \omega p$ reaction 
with linearly 
polarized photons in  horizontal (vertical) direction relative to the $\omega N$  production plane: 
\bea
\vec \xi_\parallel =  \vec\xi_x 
&=&\frac{-1}{\sqrt{2}}(\vec \xi_{+1}-\vec \xi_{-1})\nonumber\\
\vec \xi_\perp = \vec \xi_y 
&=& \frac{\mi}{\sqrt{2}}(\vec\xi_{+1}+\vec\xi_{-1}).
\eea
The coordinate system is defined by $\vec z= \vec k/|\vec k|$,  
$\vec y= \vec k\times \vec k'/|\vec k\times \vec k'|$, 
were $\vec k$($\vec k'$) is a photon(meson) three-momentum.

The following asymmetries can be also defined to test the reaction amplitude 
\bea
\Sigma_A &=& \frac{\rho^1_{11} + \rho^1_{1-1}}{\rho^0_{11} + \rho^0_{1-1}},\nonumber\\
\Sigma_B &=& \frac{\rho^1_{11} - \rho^1_{1-1}}{\rho^0_{11} - \rho^0_{1-1}}.
\label{BAasymm} \; 
\eea
The differences between $\Sigma_A$ and $\Sigma_B$ can be seen on the example of the single
$\pi^0$ exchange mechanism  (see below).
 
\subsection{Single $\pi^0$ exchange contribution to the $\omega$ photoproduction.}
\label{pi0-exchange}
In the case of the single $\pi^0$ exchange contribution  corresponding to the $(c)$ diagram in 
Fig. \refe{diag} the asymmetry and spin density matrix of the $\gamma p \to \omega p$ reaction 
can be easily calculated. The coupling for the $\omega \to \gamma \pi^0$ decay is 
proportional to 
\bea
\mcl_{\omega\gamma\pi^0} \sim \varepsilon_{\mu\nu\rho\sigma}F^{\mu\nu}\partial^\rho \omega^\sigma \pi,
\label{pi0coupling}
\eea
where $\omega^\sigma$($\pi$), and $F^{\mu\nu}$ stand for the vector meson(pion) field and the
electormagnetic tensor respectively. $\varepsilon_{\mu\nu\rho\sigma}$ is a Levi-Chivita tenzor.   
Then the amplitude of the reaction can be written as
\bea
\xi^\gamma_\nu\xi^\omega_\sigma J^{\nu\sigma} \sim  \varepsilon_{\mu\nu\rho\sigma} 
k^\mu q^\rho D_\pi(p-p')M(p,p',s_N,s_{N'}),
\label{pi0ampli}
\eea
where $p$($s_N$) and $p'$($s_{N'}$) are the four momentum(spin) of the initial and final nucleon
respectively and $k$,$q$ are the photon and vector meson four momenta. The polarization vectors of the
photon and vector meson are 
\bea
\xi^\mu_{\pm 1} =
\frac{ \mp 1}{\sqrt{2}} (\xi^\mu_x \pm \mi \xi^\mu_y),
\label{pi0p1}
\eea
for $\lambda_\gamma$,$\lambda_\omega$=$\pm 1$ and
\bea
\xi^\mu_{0} = \xi^\mu_z
\label{pi0p2}
\eea
for  longitudinally polarized $\omega$ mesons. 
From  Eqs. (\ref{rho0}-\ref{BAasymm}) it follows that only the 
current
\bea
J_{\mu\sigma}=\varepsilon_{\mu\nu\rho\sigma}k^\mu q^\rho
\label{pi0curr}
\eea
in Eq. \refe{pi0ampli} 
contributes to the spin density matrix and asymmetry.
Then the final expressions for the spin density matrices are:
\bea
\rho^0_{\lambda_\omega\lambda_{\omega'}} &=& -\frac{1}{2} 
\left(\frac{m^2_\omega(k\xi^*_{\lambda_\omega})(k\xi_{\lambda_\omega'})}
{(kq)^2} +   (\xi^*_{\lambda_\omega} \xi_{\lambda_\omega'}) \right ),\nonumber\\
\rho^1_{\lambda_\omega\lambda_{\omega'}} &=& \rho^0_{\lambda_\omega\lambda_{\omega'}} 
-\xi^{\mu*}_{\lambda_\omega}
 \xi^\nu_{\lambda_\omega'} \delta_{\mu,2}\delta_{\nu,2}.
\label{pi0rho}
\eea
From \refe{pi0curr} and \refe{pi0rho} one deduces that the single
$\pi^0$ exchange contribution in the helicity frame leads to  
$\Sigma$=0, $\Sigma_A$=-1, and   $\Sigma_B$=+1. 

In the Gottfried-Jackson frame defined as a center of mass system of the incoming photon and 
the exchange pion where the quantization axis is aligned with a photon momentum, 
the spin density matrix elements given by \refe{pi0rho} 
becomes $\rho^{0,GJ}_{11}$=$\rho^{0,GJ}_{-1-1}$=$\frac{1}{2}$ and $\rho^{0,GJ}_{00}$=0.

\section{$\omega N$ scattering lengths and effective ranges}
\label{scattering_lengths}
The spin averaged $\omega N$ scattering lengths and effective ranges are calculated using the 
convention of Lutz {\it et al} \cite{Lutz:2001}:
\bea
\bar a &=& \smallfrac{1}{3}\bar a(J=\sfoh) + \smallfrac{2}{3}\bar a(J=\sfth), \nonumber\\
\bar r &=& \smallfrac{1}{3}\bar r(J=\sfoh) + \smallfrac{2}{3}\bar r(J=\sfth).
\label{scl_definition}
\eea
The $\omega N$ helicity state combinations at threshold are \cite{Lutz:2001}
\bea
|\omega N; J= \sfoh \rangle =  |\omega N,\sfoh; J= \sfoh \rangle + \smallfrac{1}{\sqrt{2}} 
|\omega N,+0; J= \sfoh \rangle,\nonumber\\  
|\omega N; J= \sfth \rangle =  |\omega N,\sfth; J= \sfth \rangle 
+ \smallfrac{1}{\sqrt{3}} |\omega N,\sfoh; J= \sfth \rangle
+ \sqrt{\smallfrac{2}{3}} |\omega N,+0; J= \sfth \rangle.  
\label{helct_comb}
\eea

\end{appendix}
\newpage

\bibliographystyle{h-physrev3}
\bibliography{tau}

\end{document}